\documentclass[prd,onecolumn,notitlepage,amsmath,nofootinbib,superscriptaddress,showpacs,showkeys]{revtex4-1}

\usepackage[T1]{fontenc}
\usepackage[latin9]{inputenc}
\usepackage{amsmath}
\usepackage{amssymb}
\usepackage{esint}
\usepackage[brazil,english]{babel}
\usepackage{graphicx}
\usepackage[usenames,dvipsnames]{color}
\usepackage{longtable}
\usepackage{dcolumn}
\usepackage{ulem}
\usepackage{nicefrac}

\begin{document}

\title{Thermodynamics of bosonic systems in anti--de Sitter spacetime}

\author{Walace S. Elias}
\email{walace.elias@ufra.edu.br}
\affiliation{Instituto Ciberespacial, Universidade Federal Rural da Amaz\^{o}nia, Av. Presidente Tancredo Neves, 66077-901, Bel\'{e}m-PA, Brazil}

\author{C. Molina}
\email{cmolina@usp.br}
\affiliation{Escola de Artes, Ci\^{e}ncias e Humanidades, Universidade de S\~{a}o Paulo, Av. Arlindo Bettio 1000, CEP 03828-000, S\~{a}o Paulo-SP, Brazil}

\author{M. C. Baldiotti}
\email{baldiotti@uel.br}
\affiliation{Departamento de F\'{\i}sica, Universidade Estadual de Londrina, 86051-990, Londrina-PR, Brazil}

\begin{abstract}

We analyze the thermodynamics of massless bosonic systems in
$D$-dimensional anti--de Sitter spacetime, considering scalar, electromagnetic, and gravitational fields. Their dynamics are described by P\"{o}schl-Teller effective potentials and quantized in a unified framework, with the determination of the associated energy spectra.
From the microscopic description developed, a macroscopic thermodynamic treatment is proposed, where an effective volume in anti--de Sitter geometry is defined and a suitable thermodynamic limit is considered. 
Partition functions are constructed for the bosonic gases, allowing the determination of several thermodynamic quantities of interest. 
With the obtained results, general aspects of the thermodynamics are explored. 
\end{abstract}



\maketitle

\section{Introduction}

Asymptotically anti--de Sitter geometries gained a new relevance with
the anti--de Sitter/conformal field theory (AdS/CFT) correspondences
\cite{Maldacena:1997re,Witten:1998qj,Gubser:1998bc}. In the best-known
scenario, this duality establishes a dictionary between gravitational
dynamics in anti--de Sitter spacetime and $\mathcal{N}=4$ supersymmetric
Yang-Mills theory. Generalizations were proposed, considering other
geometries and field theories, leading to the gauge/gravity correspondences.
Applications in a variety of physical settings were developed, from
fundamental quantum gravity models to phenomenological condensed matter
systems.

Despite the present success of the AdS/CFT program, relevant questions
remain for the complete understanding of the duality. In particular,
there are still open issues concerning the relation between the thermodynamics
of the gravity and field theory sides. Specifically focusing on the
gravitational physics, thermodynamic aspects of asymptotically AdS
spacetimes have received considerable attention in recent years \cite{brown,louko,hemming,hubeny,denef,Rabin,menoufi,myung,cardoso,dolan,lemos,Kubiznak:2016qmn,Baldiotti:2017ywq,Bhattacharya:2017hfj}.

An important related theme is the behavior of a thermal gas in AdS spacetime, eventually collapsing to form a black hole. Hawking and Page, within a Euclidean path integral approach, considered the thermodynamics of thermal gases and black holes in AdS space \cite{Haw1983}. They observed that the Schwarzschild--anti--de Sitter (SAdS) and the thermal AdS geometries are different phases of a single physical system.

In the present work, we consider the thermal anti--de Sitter spacetime,
that is, the setup treated by Hawking-Page in the no--black hole regime.
We analyze the thermodynamics of bosonic systems in AdS geometry within
a real time approach. Scalar, electromagnetic and gravitational fields
are considered. A fundamental issue in this framework is the perturbative
and quantum dynamics in AdS background. Results concerning the decomposition
of fields of interest in AdS are commented in Refs.~\cite{Ishibashi:2003ap,Ishibashi:2003jd,Ishibashi:2004wx,Kodama:2003jz,morgan}.
The quantum scalar field in AdS spacetime was considered in \cite{Avis:1977yn,Burgess:1984ti,Cotabreveescu:1999em},
and the quantization of the electromagnetic field was studied in Refs.~\cite{Higuchi:1986ww,PhysRevD.57.1108,PhysRevD.63.124008,Elias:2014pca}.

To characterize the thermodynamics of a bosonic gas in AdS
geometry, a unified framework for the quantization of
the scalar, electromagnetic and gravitational fields is introduced. 
Energy spectra
for the physical modes are derived and partition functions
are constructed. In a suitable thermodynamic limit, quantities of
interest are calculated and an overall analysis is performed. The
present work suggests that AdS geometry qualitatively behaves as a confining box
(as expected), but with correction terms reflecting the nontrivial
geometric background. 

The structure of this paper is presented in the following. In Sec.~\ref{ads-fields} the AdS spacetime and the fields of interest are introduced. In Sec.~\ref{sec_spectra} the bosonic systems are quantized and the energy spectra determined. In Sec.~\ref{sec_thermodynamics} the thermodynamics of the considered
scenarios are analyzed. A qualitative description of the thermodynamic
characteristics and instabilities of the considered systems is
presented in Sec.~\ref{analysis}. Final remarks are made in the
Sec.~\ref{final-remarks}. 
We use signature $(-,+,\cdots,+)$ and natural units with $G=\hbar=c=k_{B}=1$
throughout this paper.

\section{anti--de Sitter spacetime and field equations}

\label{ads-fields}

\subsection{Background geometry}

\label{background}

In this work we consider the $D$-dimensional anti--de Sitter (AdS)
spacetime. More precisely, we will denote as AdS geometry the universal
covering space of the maximally symmetric Lorentzian geometry with
negative constant curvature. With this definition, AdS geometry does
not contain any closed timelike curves and it is a well-behaved solution
of vacuum Einstein equations with a negative cosmological constant
$\Lambda$.

The $D$-dimensional AdS geometry $(M,g_{\mu\nu})$ can be locally
described as a product of a two-dimensional manifold $\mathcal{M}^{2}$
and the $(D-2)$-dimensional sphere $S^{D-2}$. The subspace $\mathcal{M}^{2}$
is spanned by a timelike coordinate $t$ and a radial spacelike
coordinate $r$. The coordinate system based on $\left\{ \theta^{1},\theta^{2},\ldots,\theta^{D-2}\right\} $
describes $S^{D-2}$. With this decomposition, the background metric
can be written in the form 
\begin{equation}
ds^{2} = -f(r)\, dt^{2}+\frac{1}{f(r)}\, dr^{2}+r^{2}\,\gamma_{ij}d\theta^{i}d\theta^{j} \, ,
\label{generalsolution}
\end{equation}
where $\gamma_{ij}$ is the metric of the $(D-2)$-dimensional unit
sphere $S^{D-2}$, 
\begin{equation}
\gamma_{ij}d\theta^{i}d\theta^{j}=\left(d\theta^{1}\right)^{2}+\sin^{2}\theta^{1}\,\left(d\theta^{2}\right)^{2}+\cdots+\sin^{2}\theta^{1}\cdots\,\sin^{2}\theta^{D-2}\,\left(d\theta^{D-2}\right)^{2}\,,\label{dOmega2}
\end{equation}
and the function $f(r)$ is given by 
\begin{equation}
f(r)=1+\frac{r^{2}}{L^{2}}\,.\label{f}
\end{equation}
The parameter $L$, the ``AdS radius'', is related to the negative cosmological constant $\Lambda$ as 
\begin{equation}
L^{2}\equiv-\frac{(D-1)(D-2)}{2\Lambda}\,.
\end{equation}
The choice of the foliation in Eq.~(\ref{generalsolution}) is suitable
to the investigation of the equations of motion associated to the
bosonic fields.

A convenient chart is based on the so-called tortoise coordinate $x$, defined as
\begin{equation}
x \equiv \int\,\frac{dr}{f(r)}=L\,\arctan\,\left(\frac{r}{L}\right)\,.
\end{equation}
For the AdS geometry, a simple expression for the function $r(x)$
can be obtained, $r = L \tan (x/L)$. 
In terms of the tortoise coordinate $x$, the metric $\tilde{g}_{ab}$
on the two-dimensional AdS geometry $(\mathcal{M}^{2},\tilde{g}_{ab})$
is written as 
\begin{equation}
\tilde{g}_{ab}dy^{a}dy^{b}=\sec^{2}\left(\frac{x}{L}\right)\left(-dt^{2}+dx^{2}\right)\,,\label{metric-tortoise}
\end{equation}
with $y^{0}=t$ and $y^{1}=x$. In the following sections, the covariant
derivatives on $M$, $\mathcal{M}^{2}$ and $S^{D-2}$ are denoted
by $\nabla_{\mu}$, $\tilde{\nabla}_{a}$ and $D_{i}$, respectively.

Anti--de Sitter spacetime has interesting properties. An important feature
of this geometry is that its spatial infinity is timelike. 
Massless particles reach spatial infinity in a finite time, according
to a static observer. Assuming specific boundary conditions, usually
denoted as ``reflexive boundary conditions'' \cite{Avis:1977yn},
massive and massless particles can be confined into AdS geometry.

From those considerations, AdS spacetime can be interpreted as a box
with an effective volume $V_{\textrm{eff}}$, associated to a given
static observer. Consider an inertial observer that sends a massless
particle outward. According to this observer, the time $t_{\infty}$
for the particle to reach spatial infinity is 
\begin{equation}
t_{\infty}=\frac{\pi}{2}\, L\,.\label{effective_radius}
\end{equation}
A natural definition for an effective radius ($R_{\textrm{eff}}$) is
$R_{\textrm{eff}} \equiv c \, t_{\infty} = c\pi L/2$, where we have introduced
back the speed of light $c$ for clarity. Neglecting nonessential
proportionality factors, we define an effective volume associated
with AdS geometry as 
\begin{equation}
V_{\textrm{eff}}\equiv L^{D-1} \, .
\label{volumeeffn}
\end{equation}
The notion of an AdS effective volume will be important for the implementation
of a thermodynamic limit on the field dynamics at the anti--de Sitter
spacetime. Further comments on this definition will be presented in Sec.~\ref{volume-homogeneity}.

\subsection{Massless scalar field}

\label{scalar_field_sec}

The simplest bosonic system considered is a massless scalar field
$\Phi$ satisfying the Klein-Gordon equation, 
\begin{equation}
\Box\Phi=0\,,\label{KG-equation}
\end{equation}
where $\Box$ is the Laplace-Beltrami operator. We expand $\Phi$
in terms of scalar spherical harmonics $\mathbb{S}^{k_{S}}$ as 
\begin{equation}
\Phi(t,x,\{\theta^{i}\})=\sum_{\ell=0}^{\infty}\,\sum_{\{m_{i}\}}\,\left[L\tan\left(\frac{x}{L}\right)\right]^{-\frac{D-2}{2}}\,\phi_{\ell}^{\textrm{sc}}(t,x)\,\mathbb{S}^{k_{S}}(\{\theta^{i}\})\,.\label{expansionphi}
\end{equation}
In Eq.~(\ref{expansionphi}), $\{k_{S}^{2}\}$ are the eigenvalues
of $\mathbb{S}^{k_{S}}$ \cite{Ishibashi:2003ap}, given by 
\begin{equation}
k_{S}^{2}=\ell(\ell+D-3)\,,\,\,\ell=0,1,2,\ldots\,.\label{scalar_harmonic_eigenvalue}
\end{equation}
The integers $\{m_{i}\}$ label the modes in each subspace spanned
by the eigenmodes of $\mathbb{S}^{k_{S}}$.

The equation of motion for each mode $\phi_{\ell}^{\textrm{sc}}$
is obtained by straightforward substitution of the expansion~(\ref{expansionphi})
in Eq.~(\ref{KG-equation}), 
\begin{equation}
\frac{\partial^{2}\phi_{\ell}^{\textrm{sc}}(t,x)}{\partial t^{2}}=\left[\frac{\partial^{2}}{\partial x^{2}}-\mathcal{V}^{\textrm{sc}}(x)\right]\phi_{\ell}^{\textrm{sc}}(t,x)\,,\label{differential_equation_scalar_field}
\end{equation}
where $\mathcal{V}^{\textrm{sc}}(x)$ is a scalar effective potential,
the explicit expression of which is
\begin{equation}
\mathcal{V}^{\textrm{sc}}(x)=\frac{1}{L^{2}}\left[\frac{A^{\textrm{sc}}}{\cos^{2}\left(\frac{x}{L}\right)}+\frac{B^{\textrm{sc}}}{\sin^{2}\left(\frac{x}{L}\right)}\right]\,.\label{effective_potential}
\end{equation}
The coefficients $A^{\textrm{sc}}$ and $B^{\textrm{sc}}$ for the
massless scalar potential are 
\begin{equation}
A^{\textrm{sc}}=\frac{D(D-2)}{4}\,,\label{coeficienteA}
\end{equation}
\begin{equation}
B^{\textrm{sc}}=k_{S}^{2}+\frac{(D-2)(D-4)}{4}=\ell(\ell+D-3)+\frac{(D-2)(D-4)}{4}\,,\label{coeficienteB}
\end{equation}
with the multipole index $\ell$ assuming values in $\ell=0,1,2,\ldots\,$.

\subsection{Electromagnetic field}
\label{eletromagnetic_field_sec}

A more complex bosonic perturbation in AdS background is the electromagnetic
field. Its dynamics is determined by the electromagnetic tensor $F_{\mu\nu}$,
subjected to Maxwell's equations, 
\begin{equation}
\nabla_{\nu}F^{\mu\nu}=0\,\,,\,\nabla_{[\sigma}F_{\mu\nu]}=0\,.\label{maxwell-eq-1}
\end{equation}
From the classical electromagnetic tensor $F_{\mu\nu}$, the potential
$A_{\mu}$ is defined as 
\begin{equation}
F_{\mu\nu}=\nabla_{\mu}A_{\nu}-\nabla_{\nu}A_{\mu}\,.
\end{equation}
The equations of motion~(\ref{maxwell-eq-1}) can be decomposed into
two parts, decoupling $A_{\mu}$ into vector and scalar modes ($A_{\mu}^{\textrm{el-vc}}$
and $A_{\mu}^{\textrm{el-sc}}$, respectively). Each of these modes, after being
expanded in terms of appropriate spherical harmonics, generates a
set of second order linear differential equations, characterized by
certain effective potentials. In the following, the formalism presented
in Refs.~\cite{Ishibashi:2003ap,Ishibashi:2003jd,Ishibashi:2004wx} will
be used.

\subsubsection{Electromagnetic vector modes}

Electromagnetic vector modes $A_{\mu}^{\textrm{el-vc}}$ can be written as an expansion
in vector harmonics $\{\mathbb{V}_{i}^{k_{V}}\}$ as 
\begin{equation}
A_{\mu}^{\textrm{el-vc}}\, dx^{\mu}=\sum_{\ell=1}^{\infty}\,\sum_{\{m_{i}\}}\,\phi_{\ell}^{\textrm{el-vc}}(x)\mathbb{V}_{i}^{k_{V}}dz^{i}\,,\label{electromagnetic-modes}
\end{equation}
where $\phi_{\ell}^{\textrm{el-vc}}(x)$ is the master variable associated
to this mode. The function $\phi_{\ell}^{\textrm{el-vc}}(r)$ represents
the radial modes of $A_{\mu}^{\textrm{el-vc}}$. The vector spherical harmonics
$\mathbb{V}_{i}^{k_{V}}$ satisfy the eigenvector equation $\left(D^{j}D_{j}+k_{V}^{2}\right)\mathbb{V}_{i}^{k_{V}}=0$,
with eigenvalues $\left\{ k_{V}^{2}\right\} $ given by 
\begin{equation}
k_{V}^{2}=\ell(\ell+D-3)-1\,,\,\,\ell=1,2,\ldots\,.\label{kV}
\end{equation}

%
Using the expansion~\eqref{electromagnetic-modes}, it is obtained a differential equation associated to the master variable $\phi_{\ell}^{\textrm{el-vc}}(x)$. The new wave equation has the same form of Eq.~(\ref{differential_equation_scalar_field}), but with an effective potential $\mathcal{V}^{\textrm{el-vc}}(x)$ given by
\begin{equation}
\mathcal{V}^{\textrm{el-vc}}(x)=\frac{1}{L^{2}}\left[\frac{A^{\textrm{el-vc}}}{\cos^{2}\left(\frac{x}{L}\right)}+\frac{B^{\textrm{el-vc}}}{\sin^{2}\left(\frac{x}{L}\right)}\right]\,.
\end{equation}
The coefficients $A^{\textrm{el-vc}}$ and $B^{\textrm{el-vc}}$ are
\begin{equation}
A^{\textrm{el-vc}}=\frac{(D-2)(D-4)}{4}\,,
\end{equation}
\begin{equation}
B^{\textrm{el-vc}}=k_{V}^{2}+1+\frac{(D-4)^{2}}{4}=\ell(\ell+D-3)+\frac{(D-2)(D-4)}{4}\,,
\end{equation}
with the multipole index $\ell=1,2,\ldots$.

\subsubsection{Electromagnetic scalar modes}

Electromagnetic scalar modes $A_{\mu}^{\textrm{el-sc}}$ can be written in terms
of two quantities, a vector $A_{a}$ orthogonal to $S^{n}$ and a
scalar quantity $A$. Hence, with a convenient decomposition in spherical
harmonics, $A_{\mu}^{\textrm{el-sc}}$ can be expressed as 
\begin{equation}
A_{\mu}^{\textrm{el-sc}}dx^{\mu}=\sum_{\ell=1}^{\infty}\,\sum_{\{m_{i}\}}\,\left(A_{\ell a}\mathbb{S}^{k_{S}}dy^{a}+A_{\ell}D_{i}\mathbb{S}^{k_{S}}d\theta^{i}\right)\,.\label{eq22}
\end{equation}

From the first Maxwell equation in~(\ref{maxwell-eq-1}), one obtains
that 
\begin{equation}
\tilde{\nabla}_{a}\left\{ \left[L\tan\left(\frac{x}{L}\right)\right]^{D-4}(\tilde{\nabla}^{a}A_{\ell}+k_{S}A_{\ell}^{a})\right\} =0\,,\label{eq26}
\end{equation}
where the quantity $k_{S}^{2}$ is the scalar harmonic eigenvalue
presented in Eq.~(\ref{scalar_harmonic_eigenvalue}). The result
in Eq.~(\ref{eq26}) implies \cite{Ishibashi:2003ap} that there
is a function $\phi_{\ell}^{\textrm{el-sc}}$ satisfying
\begin{equation}
\tilde{\nabla}_{a}\phi_{\ell}^{\textrm{el-sc}}=g_{ab}\,\left[L\tan\left(\frac{x}{L}\right)\right]^{D-4}(\tilde{\nabla}^{b}A_{\ell}+k_{S}A_{\ell}^{b})\,.\label{eq27}
\end{equation}
A master variable $\tilde{\phi}_{\ell}^{\textrm{el-sc}}$ is then introduced,
\begin{equation}
\tilde{\phi}_{\ell}^{\textrm{el-sc}}\equiv\left[L\tan\left(\frac{x}{L}\right)\right]^{-\frac{D-4}{2}}\phi_{\ell}^{\textrm{el-sc}}\label{phi-el-sc}
\end{equation}
and, in terms of $\tilde{\phi}_{\ell}^{\textrm{el-sc}}$, electromagnetic
scalar modes are characterized by a wave equation in the form~(\ref{differential_equation_scalar_field})
with the effective potential \cite{Ishibashi:2003ap} 
\begin{equation}
\mathcal{V}^{\textrm{el-sc}}=\frac{1}{L^{2}}\left[\frac{A^{\textrm{el-sc}}}{\cos^{2}\left(\frac{x}{L}\right)}+\frac{B^{\textrm{el-sc}}}{\sin^{2}\left(\frac{x}{L}\right)}\right]\,.\label{potencialescalar}
\end{equation}
The constants $A^{\textrm{el-sc}}$ and $B^{\textrm{el-sc}}$ are
given by: 
\begin{equation}
A^{\textrm{el-sc}}=\frac{(D-4)(D-6)}{4}\,,\label{A-el-sc}
\end{equation}
\begin{equation}
B^{\textrm{el-sc}}=k_{S}^{2}+\frac{(D-2)(D-4)}{4}=\ell(\ell+D-3)+\frac{(D-2)(D-4)}{4}\,,\label{B-el-sc}
\end{equation}
with the multipole index $\ell=1,2,\ldots$.

\subsection{Gravitational perturbation}
\label{gravitational_perturbations_sec}

Quantum properties of the gravitational field can be considered in
a field theory effective approach. Gravitational perturbations propagating
in AdS spacetime can be expanded in terms of harmonic functions on
$S^{D-2}$. The perturbed Einstein equations are then expressed in
terms of a set of gauge invariant quantities \cite{Kodama:2003jz}.
These quantities are combinations of the metric perturbations $h_{\mu\nu}$,
which are related to the perturbed spacetime metric $g_{\mu\nu}$
as 
\begin{equation}
g_{\mu\nu}=g_{\mu\nu}^{(0)}+h_{\mu\nu}\,,
\end{equation}
with $g_{\mu\nu}^{(0)}$ representing the AdS metric. By taking appropriate
combinations of gauge invariant variables constructed from $h_{\mu\nu}$,
the perturbative equations are reduced to three decoupled sets. They
furnish the tensor, vector and scalar perturbations, which will
be described in the following with the formalism of Refs.~\cite{Ishibashi:2003ap,Ishibashi:2003jd,Ishibashi:2004wx}.

\subsubsection{Gravitational tensor modes}

Gravitational tensor perturbations are present in spacetimes with
dimension equal to or larger than $5$. This set of gravitational modes
can be represented in terms of tensor spherical harmonics as follows, 
\begin{equation}
h_{ab} = 0 \,, \,\,\,
h_{ai}=0 \, , \,\,\,
h_{ij}=\sum_{\ell=2}^{\infty}\,\sum_{\{m_{i}\}}\,\frac{2}{L^{2}\tan^{2}\left(\frac{x}{L}\right)}\, H_{\ell}^{T}(t,x)\mathbb{T}_{ij}^{k_{T}}\,,
\label{eq39}
\end{equation}
where $\mathbb{T}_{ij}^{k_{T}}$ are transverse traceless harmonic
tensors on $S^{D-2}$, with the quantities $\{k_{T}^{2}\}$ defined as
\begin{equation}
k_{T}^{2}=\ell(\ell+D-3)-2\,,\,\,\ell=2,3,\ldots\,.
\end{equation}
Using the master variable \cite{Ishibashi:2003ap} given by 
\begin{equation}
\phi_{\ell}^{\textrm{gr-t}}\equiv\left[L\tan\left(\frac{x}{L}\right)\right]^{\frac{D-6}{2}}H_{\ell}^{T}\,,
\end{equation}
the equation of motion for the gravitational tensor mode assumes the
form of Eq.~(\ref{differential_equation_scalar_field}), with an
effective potential $\mathcal{V}^{\textrm{gr-t}}$ written as 
\begin{equation}
\mathcal{V}^{\textrm{gr-t}}=\frac{1}{L^{2}}\left[\frac{A^{\textrm{gr-t}}}{\cos^{2}\left(\frac{x}{L}\right)}+\frac{B^{\textrm{gr-t}}}{\sin^{2}\left(\frac{x}{L}\right)}\right]\,.
\end{equation}
The coefficients $A^{\textrm{gr-t}}$ and $B^{\textrm{gr-t}}$ are
\begin{equation}
A^{\textrm{gr-t}}=\frac{D(D-2)}{4}\,,
\end{equation}
\begin{equation}
B^{\textrm{gr-t}}=k_{T}^{2}+2+\frac{(D-2)(D-4)}{4}=\ell(\ell+D-3)+\frac{(D-2)(D-4)}{4}\,,\label{B-gr-t}
\end{equation}
with the multipole index $\ell=2,3,\ldots$.

\subsubsection{Gravitational vector modes}

Gravitational vector modes can be expanded in terms of vector harmonic
functions $\mathbb{V}_{i}^{k_{V}}$ as 
\begin{equation}
h_{ab}=0\,,\,\, h_{ai}=\sum_{\ell=2}^{\infty}\,\sum_{\{m_{i}\}}\,\frac{1}{L\,\tan\left(\frac{x}{L}\right)}f_{\ell a}(t,x)\,\mathbb{V}_{i}^{k_{V}}\, , \,\, 
h_{ij}=\sum_{\ell=2}^{\infty}\,\sum_{\{m_{i}\}}\,\frac{2}{L^{2}\tan^{2}\left(\frac{x}{L}\right)}\, H_{\ell}^{V}(t,x)\mathbb{V}_{ij}^{k_{V}}\,.
\end{equation}
%
%
The quantities $\mathbb{V}_{ij}^{k_{V}}$ are vector-type harmonic
tensors on $S^{D-2}$ built from the transverse harmonic vectors $\mathbb{V}_{i}^{k_{V}}$,
with the eigenvalue $k_{V}^{2}$ presented in Eq.~(\ref{kV}). From
the functions $f_{\ell a}$ and $H_{\ell}^{V}$, a new gauge invariant
quantity $Z_{\ell a}$ is defined: 
\begin{equation}
Z_{\ell a}\equiv f_{\ell a}(t,x)-L^{2}\tan^{2}\left(\frac{x}{L}\right)\tilde{\nabla}_{a}\left[\frac{H_{\ell}^{V}(t,x)}{\tan\left(\frac{x}{L}\right)^{2}}\right]\,.\label{eq51}
\end{equation}
A master variable $\phi_{\ell}^{\textrm{gr-vc}}$ is implicitly introduced
by the relation 
\begin{equation}
Z_{\ell a}=\left[L\tan\left(\frac{x}{L}\right)\right]^{-(D-4)}\epsilon_{ab}\tilde{\nabla}^{b}\phi_{\ell}^{\textrm{gr-vc}}\,,\label{eq54}
\end{equation}
where $\epsilon_{ab}$ is the Levi-Civita tensor in 
$\mathcal{M}^{2}$. In terms of $\phi_{\ell}^{\textrm{gr-vc}}$, vector-type gravitational modes satisfy a wave equation with the form of Eq.~(\ref{differential_equation_scalar_field})~\cite{Ishibashi:2003jd},
where the effective potential $\mathcal{V}^{\textrm{gr-vc}}$ is 
\begin{equation}
\mathcal{V}^{\textrm{gr-vc}}=\frac{1}{L^{2}}\left[\frac{A^{\textrm{gr-vc}}}{\cos^{2}\left(\frac{x}{L}\right)}+\frac{B^{\textrm{gr-vc}}}{\sin^{2}\left(\frac{x}{L}\right)}\right]\,.
\end{equation}
The constants $A^{\textrm{gr-vc}}$ and $B^{\textrm{gr-vc}}$ are
given by 
\begin{equation}
A^{\textrm{gr-vc}}=\frac{(D-2)(D-4)}{4}\,,
\end{equation}
\begin{equation}
B^{\textrm{gr-vc}}=k_{V}^{2}+1+\frac{(D-2)(D-2)}{4}=\ell(\ell+D-3)+\frac{(D-2)(D-4)}{4}\,,\label{B-gr-vc}
\end{equation}
with the multipole index assuming values in $\ell=2,3,\ldots$.

\subsubsection{Gravitational scalar modes}

Gravitational scalar perturbations can be expanded in terms of scalar
harmonic functions $\{\mathbb{S}^{k_{S}}\}$ as 
\begin{gather}
h_{ab}=\sum_{\ell=2}^{\infty}\,\sum_{\{m_{i}\}}\, f_{\ell ab}(t,x)\mathbb{S}^{k_{S}}\,,\,\, h_{ai}=\sum_{\ell=2}^{\infty}\,\sum_{\{m_{i}\}}\,\frac{1}{L\,\tan\left(\frac{x}{L}\right)}\, f_{\ell a}(t,x)\mathbb{S}_{i}^{k_{S}}\,,\nonumber \\
h_{ij}=\sum_{\ell=2}^{\infty}\,\sum_{\{m_{i}\}}\,\frac{2}{L^{2}\tan^{2}\left(\frac{x}{L}\right)}\,\left[H_{\ell}^{L}(t,x)\gamma_{ij}\mathbb{S}^{k_{S}}+H_{\ell}^{S}(t,x)\mathbb{S}_{ij}^{k_{S}}\right]\,.
\end{gather}
The terms $\mathbb{S}_{i}^{k_{S}}$ and $\mathbb{S}_{ij}^{k_{S}}$
are scalar-type harmonic vectors and tensors on $S^{D-2}$, respectively,
built from the scalar harmonic functions $\mathbb{S}^{k_{S}}$ as
\begin{equation}
\mathbb{S}_{i}^{k_{S}}=-\frac{1}{k_{S}}D_{i}\mathbb{S}^{k_{S}}\,,\,\,\mathbb{S}_{ij}^{k_{S}}=\frac{1}{k_{S}^{2}}D_{i}D_{j}\mathbb{S}^{k_{S}}+\frac{1}{2}\gamma_{ij}\mathbb{S}^{k_{S}}\,.
\end{equation}

Gauge invariant quantities can be defined for $\ell\geq2$ and written
in terms of a master variable $\phi_{\ell}^{\textrm{gr-sc}}$ (after
a rather involving procedure, as seen for example in Ref.~\cite{Ishibashi:2004wx}).
The function $\phi_{\ell}^{\textrm{gr-sc}}$ satisfies a wave equation
with the form of Eq.~(\ref{differential_equation_scalar_field})
with an effective potential given by $\mathcal{V}^{\textrm{gr-sc}}$,
where 
\begin{equation}
\mathcal{V}^{\textrm{gr-sc}}=\frac{1}{L^{2}}\left[\frac{A^{\textrm{gr-sc}}}{\cos^{2}\left(\frac{x}{L}\right)}+\frac{B^{\textrm{gr-sc}}}{\sin^{2}\left(\frac{x}{L}\right)}\right]\,.
\end{equation}
The coefficients $A^{\textrm{gr-sc}}$ and $B^{\textrm{gr-sc}}$ are
given by 
\begin{equation}
A^{\textrm{gr-sc}}=\frac{(D-4)(D-6)}{4}\,,\label{A-gr-sc}
\end{equation}
\begin{equation}
B^{\textrm{gr-sc}}=k_{S}^{2}+\frac{(D-2)(D-4)}{4}=\ell(\ell+D-3)+\frac{(D-2)(D-4)}{4}\,,\label{B-gr-sc}
\end{equation}
with the multipole index $\ell=2,3,\ldots$.

\section{Quantization, energy spectra and degeneracies}
\label{sec_spectra}

As seen in Sec.~\ref{ads-fields}, for the bosonic fields considered,
the classical dynamics are described by equations of motion having
the form 
\begin{equation}
\frac{\partial^{2}\phi(t,x)}{\partial t^{2}}=\left[\frac{\partial^{2}}{\partial x^{2}}-\mathcal{V}\left(x\right)\right]\phi(t,x)\,,\label{equation-general}
\end{equation}
with a P\"{o}schl-Teller~\cite{Poschl:1933zz} effective potential $\mathcal{V}(x)$,
\begin{equation}
\mathcal{V}(x)=\frac{1}{L^{2}}\left[\frac{A}{\cos^{2}\left(\frac{x}{L}\right)}+\frac{B}{\sin^{2}\left(\frac{x}{L}\right)}\right]\,.
\end{equation}
The constants $A$ and $B$ depend on the specific mode considered,
according to previous section.

In the approach used for the quantization of the scalar, electromagnetic
and gravitational perturbations in AdS spacetime, we consider the
quantum properties of the potential associated to each mode. The first
step in the quantum treatment is to define the one-particle Hilbert
space $\mathcal{H}_{1}$ associated to a given perturbation. We start
with a natural domain of functions that are absolutely continuous
in the interval $[0,\frac{\pi L}{2}]$, together with their first derivatives.
The one-particle Hamiltonian can be introduced as 
\begin{equation}
\hat{H}=\mathrm{i}\,\frac{\partial}{\partial t}\,.\label{H_explicit}
\end{equation}
That is, the notion of energy is being defined by static observers
following integral curves of $\partial/\partial t$.

From the one-particle sector, the Fock space associated with the perturbations
is constructed through the usual procedure. Taking into account Eq.~(\ref{H_explicit}),
the positive and negative energy modes 
\begin{equation}
\tilde{\phi}(x)\, e^{-\mathrm{i}\epsilon t}\,,\,\,\tilde{\phi}(x)\, e^{+\mathrm{i}\epsilon t}\,,
\end{equation}
should span a dense subset of $\mathcal{H}_{1}$, where $\tilde{\phi}(x)$
are the solutions of the ``time-independent'' version of the equation
of motion~(\ref{equation-general}): 
\begin{equation}
\hat{O}\tilde{\phi}\left(x\right) =
-\epsilon^{2}\tilde{\phi}\left(x\right) \, ,\ \, 
\hat{O}=\frac{d^{2}}{dx^{2}}-\mathcal{V}(x) \, .
\label{time-independent}
\end{equation}

An important point in the present development is that the deficiency
indices of the operator $\hat{O}$ in Eq.~(\ref{time-independent}) are not zero for arbitrary values of $A$ and $B$ \cite{Ishibashi:2004wx,Gitman}. 
Indeed, only for $A \ge 3/4$ and $B \ge 3/4$ the deficiency indices are zero, implying that the operator $\hat{O}$ is essentially self-adjoint. In this case, there is no freedom in the choice of the boundary conditions. On the other hand, if $B\ge 3/4$ and $A<3/4$, the deficiency indices are equal to $1$, implying the existence of an uniparametric family of self-adjoint extensions. In this case, which is relevant for the present work, there is more freedom in the choice of boundary conditions. 
Therefore, additional conditions must be supplied in order to specify an adjoint
extension of $\hat{O}$. This issue is related with the fact that,
for the quantum treatment of fields in AdS spacetime, the geometry
is not globally hyperbolic, with its spatial infinity being timelike.
As already commented on in Refs.~\cite{Avis:1977yn,Aharony:2010ay,Marolf:2006nd},
additional conditions must be supplied for the well-posedness of the
quantization process.
If $A<-1/4$, $\hat{O}$ is no longer positive definite, and its spectrum is unbounded from below. None of the perturbations considered here is in this class of effective potentials.
Finally, for $B < 3/4$ and $A<3/4$ the deficiency indices are equal to $2$, and the extensions are specified by two parameters and identified with elements of the group $U(2)$. This range of $A$ and $B$ is also not relevant here.

Let us initially consider the case where $A\geq3/4$ and $B\geq3/4$.
The general solution of the time-independent equation~(\ref{time-independent})
with the P\"{o}schl-Teller potential is well known (see for example Ref.~\cite{Gitman}).
It can be written as 
\begin{equation}
\tilde{\phi}(x)=C_{1}\, u_{1}(x)+C_{2}\, u_{2}(x)\,,\label{general-solution}
\end{equation}
where $u_{1}(x)$ and $u_{2}(x)$ are linearly independent solutions
of Eq.~(\ref{time-independent}) given by 
\begin{align}
u_{1}(x) & =\sin^{2p}\left(\frac{x}{L}\right)\cos^{2q}\left(\frac{x}{L}\right)F\left[\zeta_{p,q}^{\epsilon}-\frac{1}{2},\zeta_{p,q}^{-\epsilon}-\frac{1}{2};2p+\frac{1}{2};\sin^{2}\left(\frac{x}{L}\right)\right]\,,\label{u1}\\
u_{2}(x) & =\sin^{1-2p}\left(\frac{x}{L}\right)\cos^{2q}\left(\frac{x}{L}\right)F\left[\zeta_{-p,q}^{\epsilon},\zeta_{-p,q}^{-\epsilon};\frac{3}{2}-2p;\sin^{2}\left(\frac{x}{L}\right)\right]\,,\label{u2}
\end{align}
with 
\begin{equation}
\zeta_{p,q}^{\epsilon}=p+q+\frac{L\epsilon}{2}+\frac{1}{2}\,.
\end{equation}
In Eqs.~(\ref{u1}) and (\ref{u2}), $F[\alpha,\beta;\gamma;z]$
denotes the Gaussian hypergeometric function. The parameters $p$
and $q$ are expressed in terms of the constants $A$ and $B$ as
\begin{equation}
p=\frac{1}{4}+\frac{1}{2}\sqrt{B+\frac{1}{4}}\,,\,\, q=\frac{1}{4}+\frac{1}{2}\sqrt{A+\frac{1}{4}}\,.\label{p_and_q}
\end{equation}

Constants $C_{1}$ and $C_{2}$ will be fixed by the boundary conditions.
In the case $A\geq3/4$ and $B\geq3/4$, the function $u_{2}$ is
not square integrable and we must set $C_{2}=0$ in Eq.~(\ref{general-solution}).
The asymptotic behavior of $u_{1}$ at spatial infinity is \cite{Gitman}
\begin{equation}
\lim_{x\rightarrow\pi L/2}u_{1}\rightarrow\frac{\Gamma\left(2p+\frac{1}{2}\right)\Gamma\left(2q-\frac{1}{2}\right)}{\Gamma\left(p+q+\frac{L\epsilon}{2}\right)\Gamma\left(p+q-\frac{L\epsilon}{2}\right)}\left(\frac{\pi L}{2}-\frac{x}{L}\right)^{1-2q}\,.
\end{equation}
Since $\epsilon>0$, $2p+1/2>0$ and $2q+1/2>0$, for $q\geq3/4$
the function $u_{1}$ is square integrable only for specific values
of $\epsilon$ which satisfy \cite{Ishibashi:2004wx,Gitman}
\begin{equation}
p+q-\frac{L\epsilon}{2}=-n\,,\,\,\textrm{with}\,\, n=0,1,\ldots\,.\label{dirichlet-condition}
\end{equation}
Condition~(\ref{dirichlet-condition}) will give the spectra of the
several bosonic perturbations considered.

The above development is valid in the range $A\geq3/4$ and $B\geq3/4$.
However, the inequality $A\geq3/4$ does not include the electromagnetic
and gravitational scalar perturbations with $D=5$ ($A=-1/4$) or
$D=6$ ($A=0$).

For $A<3/4$, there exists an uniparametric family of self-adjoint
extensions of the operator $\hat{O}$. The determination of this family%
\footnote{In what follows, in order to use the results of Ref.~\cite{Gitman}, it
is necessary to make $x\rightarrow\pi L/2-x$, which implies $A\leftrightarrow B$.}%
depends on the asymptotic behavior of the functions $\psi_{\ast}\in\mathbb{D}(\hat{O}^{+})$
in the domain of the adjoint operator $\hat{O}^{+}$. For $A<3/4$,
we have as $x\rightarrow\pi L/2$ that~\cite{Gitman} 
\begin{equation}
\lim_{x\rightarrow\pi L/2}\psi_{\ast}\rightarrow a_{1}u_{+}\left(x\right)+a_{2}u_{-}\left(x\right)\,,\ u_{\pm}\left(x\right)=\left(\frac{\pi L}{2}-\frac{x}{L}\right)^{\pm\left(2q-\frac{1}{2}\right)+\frac{1}{2}}\,.
\end{equation}
The functions $u_{\pm}$ are square integrable for $-1\leq4A<3$.
However, in general, the asymptotic forms of $u_{\pm}$, denoted here
by $\psi_{\ast}$, are not compatible with a symmetric $\hat{O}$.
For $A\neq-1/4$, the requirement that this symmetry is enforced implies
$\bar{a}_{1}a_{2}=\bar{a}_{2}a_{1}$. This last condition can be parametrized
as~\cite{Gitman} 
\begin{equation}
a_{1}\cos\theta=a_{2}\sin\theta\,,\ \theta\in\left[-\frac{\pi}{2},\frac{\pi}{2}\right]\,.\label{theta}
\end{equation}

The variable $\theta$ introduced in Eq.~(\ref{theta}) parametrizes
the boundary conditions. Consequently $\theta$ determines the self-adjoint
extensions $\hat{O}_{\theta}$ of $\hat{O}$ for $-1/4<A<3/4$. A
physical interpretation is that $\theta$ is a parameter which controls
the amount of particles that can ``leak out'' at spatial infinity.
Indeed, the case $\theta=\pm\pi/2$ ($a_{2}=0$) fixes $\psi_{\ast}\left(\pi L/2\right)=0$.
This condition can be interpreted as a generalized Dirichlet condition~\cite{Ishibashi:2004wx}
and represents the reflexive condition at infinity. On the other hand,
$\theta=0$ ($a_{1}=0$) can be interpreted as a generalized Neumann
condition and represents a scenario where spatial infinity is transparent.

The spectrum $\epsilon_{\theta}$ of the operator $\hat{O}_{\theta}$
also depends on $\theta$. For $\theta=\pm\pi/2$, it follows that
$\epsilon_{\pm\pi/2}=\epsilon$~\cite{Gitman}. That is, the expression~(\ref{dirichlet-condition})
is obtained again. The case $A=-1/4$ shows some peculiarities but
can be treated in an analogous manner. The important point here is
that the boundary condition $\psi_{\ast}\left(\pi L/2\right)=0$ implies
that the operator $\hat{O}$ is self-adjoint. 

%
Hence, the one-particle Hilbert space $\mathcal{H}_{1}$ is the set of functions defined on the dense domain $\mathbb{D}\left(\mathbb{R}_{+}\right)$ of smooth functions with compact support which satisfy reflexive boundary conditions:
\begin{equation}
\mathbb{D}\left(\mathbb{R}_{+}\right)=\left\{ \tilde{\phi}\left(x\right):\tilde{\phi}\in C^{\infty}(\mathbb{R}_{+})\,,\,\,\mathrm{supp}\,\tilde{\phi}\subseteq[0,\frac{\pi L}{2}]\,,\,\,\tilde{\phi}(0)=\tilde{\phi}\left(\frac{\pi L}{2}\right)=0\right\} \,.
\label{novo1*}
\end{equation}
%
%
In this way, the spectrum of the operator $\hat{O}$ is given by Eq.~(\ref{dirichlet-condition}) for any values of $A$ and $B$ relevant in the present work.

With the results obtained for the P\"{o}schl-Teller potential, the spectra
of the several perturbations considered can be derived. Relation~(\ref{dirichlet-condition}),
combined with definitions~(\ref{p_and_q}) for $p$ and $q$, plus the
specific forms of the constants $A$ and $B$, determine the spectra
\begin{equation}
\epsilon^{\texttt{mode}}=\frac{1}{L}\left(2n+\ell^{\texttt{mode}}+D^{\texttt{mode}}-1\right)\,\,\,\textrm{with}\,\, n=0,1,\ldots \,,
\label{general-spectra}
\end{equation}
where $\texttt{mode}$ is a label indicating a particular perturbation,
with 
$\texttt{mode}\in\{\textrm{sc},\textrm{el-vc},\textrm{el-sc},\textrm{gr-t},\textrm{gr-vc},\textrm{gr-sc}\}$,
and 
\begin{align}
D^{\textrm{sc}} & =D^{\textrm{gr-t}}=D\,;\, D^{\textrm{el-sc}}=D^{\textrm{gr-sc}}=\frac{D+|D-5|-3}{2}\,;\, D^{\textrm{el-vc}}=D^{\textrm{gr-vc}}=D-3\,;\nonumber \\
\ell^{\textrm{sc}} & =0,1,2\ldots\,;\,\ell^{\textrm{el-vc}}=\ell^{\textrm{el-sc}}=1,2,\ldots\,;\,\ell^{\textrm{gr-t}}=\ell^{\textrm{gr-vc}}=\ell^{\textrm{gr-sc}}=2,3,\ldots\,.
\end{align}
For clarity's sake, we also present the explicit form of the spectrum
associated to each considered mode in Table~\ref{spectrum_modes_table1}.

\begin{table}[t]
\caption{Energy spectra for massless bosonic fields in AdS geometry with $D \geq4 $ and $n=0,1,2,\ldots$.}
\centering 
\begin{tabular*}{1\columnwidth}{@{\extracolsep{\fill}}@{\extracolsep{\fill}}@{\extracolsep{\fill}}c@{\extracolsep{\fill}}c@{\extracolsep{\fill}}c@{\extracolsep{\fill}}c@{\extracolsep{\fill}}}
\hline 
\textbf{Field}  & \textbf{Scalar modes}  & \textbf{Vector modes}  & \textbf{Tensor modes} \\
\hline 
\parbox[c]{0.2\textwidth}{Scalar}  &  &  & \\
($\ell=0,1,\ldots$)  & $\epsilon^{\textrm{sc}}=\frac{2n+\ell+D-1}{L}$  &  & \\
\parbox[c]{0.2\textwidth}{Electromagnetic}  &  &  & \\
($\ell=1,2,\ldots$)  & $\epsilon^{\textrm{el-sc}}=\frac{1}{L}\,\left(2n+\ell+\frac{D-1+|D-5|}{2}\right)$  & $\epsilon^{\textrm{el-vc}}=\frac{2n+\ell+D-2}{L}$  & \\
\parbox[c]{0.2\textwidth}{Gravitational}  &  &  & \\
($\ell=2,3,\ldots$)  & $\epsilon^{\textrm{gr-sc}}=\frac{1}{L}\,\left(2n+\ell+\frac{D-1+|D-5|}{2}\right)$  & $\epsilon^{\textrm{gr-vc}}=\frac{2n+\ell+D-2}{L}$  & $\epsilon^{\textrm{gr-t}}=\frac{2n+\ell+D-1}{L}$ \\
\hline 
\end{tabular*}
\label{spectrum_modes_table1} 
\end{table}

From Eq.~(\ref{general-spectra}) (or Table~\ref{spectrum_modes_table1}), we define the spectrum sequences 
\begin{equation}
\epsilon_{j}^{\texttt{mode}}:\mathbb{N}\rightarrow\{\epsilon^{\texttt{mode}}\}\,,
\end{equation}
in such a way that $\epsilon_{j}^{\texttt{mode}}<\epsilon_{j+1}^{\texttt{mode}}$.
Hence, the index $j$ labels the energy levels. 
With the energy spectra determined, their associated degeneracies
can be considered. Indeed, for a given perturbation, there are in general
several modes associated to a specific energy value $\epsilon_{j}^{\texttt{mode}}$.
We denote the degeneracy of the energy $\epsilon_{j}^{\texttt{mode}}$ as 
$\mathcal{D}_{j}^{\texttt{mode}}$.
The functions $\mathcal{D}^{\texttt{mode}}(L\epsilon_{j}^{\texttt{mode}})$ are introduced so the degeneracy of the energy level $j$ of the indicated mode ($\mathcal{D}_{j}^{\texttt{mode}}$) is written as
\begin{equation}
\mathcal{D}_{j}^{\texttt{mode}} = 
\mathcal{D}^{\texttt{mode}}(L\epsilon_{j}^{\texttt{mode}}) \, .
\label{degeneracy-function}
\end{equation}

All the considered spectra can be written as
\begin{equation}
2n + \ell = \alpha^{\textrm{mode}} \, ,
\end{equation}
where $\alpha^{\textrm{mode}}$ are obtained from Eq.~(\ref{general-spectra}) (or Table~\ref{spectrum_modes_table1}). These constants do not depend on $n$ nor $\ell$. The degeneracies will be given by the multiplicity of the eigenvalues of the spherical harmonic modes. Following the development in Ref.~\cite{Cotabreveescu:1999em},
\begin{equation}
\mathcal{D}^{\textrm{mode}} (L\epsilon) =
\frac{\Gamma\left(\alpha^{\textrm{mode}}+D-1\right)}{\left(D-2\right)! 
\, \Gamma\left(\alpha^{\textrm{mode}} + 1 \right)} \, ,
\label{degeneracies-general}
\end{equation}
with $\Gamma$ denoting the usual gamma function.
For the scalar and gravitational tensor modes, the degeneracy functions are
\begin{equation}
\mathcal{D}^{\textrm{sc}} (L\epsilon) =
\mathcal{D}^{\texttt{gr-t}}(L\epsilon) =
\frac{\Gamma(L\epsilon)}{(D-2)!\,\Gamma(L\epsilon-D+2)} \, .
\label{degenerescencia12-1}
\end{equation}
Considering the electromagnetic vector and gravitational vector modes,
\begin{equation}
\mathcal{D}^{\textrm{el-vc}}(L\epsilon) =
\mathcal{D}^{\texttt{gr-vc}}(L\epsilon) =
\frac{\Gamma(L\epsilon+1)}{(D-2)!\,\Gamma(L\epsilon-D+3)} \, .
\label{degenerescencia_eletro_vector}
\end{equation}
Finally, for the electromagnetic scalar and gravitational scalar modes,
\begin{equation}
\mathcal{D}^{\textrm{el-sc}}(L\epsilon) = 
\mathcal{D}^{\texttt{gr-sc}}(L\epsilon) =
\frac{\Gamma\left(L\epsilon-\frac{D-1+|D-5|}{2}+D-1\right)}{(D-2)!\,\Gamma\left(L\epsilon-\frac{D-1+|D-5|}{2}+1\right)} \, .
\label{degenerescencia_eletro_scalar}
\end{equation}

\section{Thermodynamics of bosonic systems}
\label{sec_thermodynamics}

\subsection{Thermodynamic and quasithermodynamic limits}

In the treatment introduced in Sec.~\ref{sec_spectra}, the fields
were assumed to be free, not allowing direct coupling among the quantum
particles. The AdS geometry has a confining potential that acts as
a box, with an effective volume $V_{\textrm{eff}}$ defined in Eq.~(\ref{volumeeffn}). We consider the anti--de Sitter geometry populated by a thermal gas, composed by massless particles with a well-defined energy and subjected to the Bose-Einstein statistics. The number of particles is not conserved, and the chemical potential $\mu$ is constant and null.

Since the bosonic system is in thermal equilibrium with
fixed temperature $T$ and fixed chemical potential ($\mu=0$), 
a grand canonical ensemble is used. 
For a given mode, the associated partition function $Z^{\texttt{mode}}$ is 
\begin{equation}
\ln Z^{\texttt{mode}}=-\sum_{j}\,\mathcal{D}_{j}^{\texttt{mode}}\,\ln\left[1-\exp\left(\frac{-\epsilon_{j}^{\texttt{mode}}}{T}\right)\right] \,,
\label{partion_function_ads}
\end{equation}
where $j$ labels the energy levels and $\mathcal{D}_{j}^{\texttt{mode}}$
is the associated degeneracy, according to expressions~(\ref{degenerescencia12-1})--(\ref{degenerescencia_eletro_scalar}).
The partition function of a given field is obtained by summing the partition functions of the modes related to that field, as will be explicitly done in the next subsections. 
A thermodynamic quantity $X$ linked to a certain bosonic field will be denoted as $X^{\texttt{field}}$, where 
$\texttt{field} \in \{ \textrm{sc}, \textrm{el}, \textrm{gr} \}$.

Macroscopic quantities can be established if it is possible to define
a proper thermodynamic limit. Using the effective volume introduced
in Eq.~(\ref{volumeeffn}) and considering that we are employing
a grand canonical ensemble, we define the thermodynamic limit \cite{Kuzemsky:2014jna} as
\begin{equation}
V_{\textrm{eff}}\rightarrow\infty\,\,\textrm{with}\,\, T\,\,\textrm{fixed}\,\,\textrm{and}\,\,\mu=0\, \,,
\label{thermodynamic_limit}
\end{equation}
which implies $L\rightarrow\infty$ with $T$ fixed and $\mu=0$.

In order to obtain the thermodynamic limit of the partition functions
in Eq.~(\ref{partion_function_ads}), we consider the norm of the
sequence $\left(\epsilon_{j}^{\texttt{mode}}:j=0,1,\ldots\right)$
for the bosonic perturbations. For the spectra in Eq.~(\ref{general-spectra}),
the associated sequence norms have the same value $\Delta\epsilon$,
\begin{equation}
\Delta\epsilon\equiv\min\left(\epsilon_{j+1}^{\texttt{mode}}-\epsilon_{j}^{\texttt{mode}}\right)=\frac{1}{L} \, .
\label{norm}
\end{equation}
Result~(\ref{norm}) indicates that a continuum limit for the sequence
can be obtained in the limit $L\rightarrow\infty$. From the expression
of the partition function, its continuum limit is derived by rewriting
Eq.~(\ref{partion_function_ads}) as a Riemann sum in the form 
\begin{equation}
\ln Z^{\texttt{mode}} = -L^{D-1} \, \sum_{j} \, \frac{\mathcal{D}_{j}^{\texttt{mode}}}{L^{D-2}}\,\ln\left[1-\exp\left(-\frac{\epsilon_{j}^{\texttt{mode}}}{T}\right)\right]\,\Delta\epsilon\,.
\end{equation}
In the thermodynamic limit, $\Delta\epsilon\rightarrow0$
and  $\mathcal{D}^{\texttt{mode}}(L\epsilon)\sim(L\epsilon)^{D-2}$
 for all perturbations discussed. Since 
\linebreak
$\mathcal{D}^{\texttt{mode}}(L\epsilon)\,\ln\left[1-\exp\left(-\frac{\epsilon}{T}\right)\right]/L^{D-2}$
is a bounded continuous function on $\epsilon$, the Riemann sum is
approximated by a Riemann integral, 
\begin{equation}
\lim_{L\rightarrow\infty}\sum_{j}\,\frac{\mathcal{D}_{j}^{\texttt{mode}}}{L^{D-2}}\,\ln\left[1-\exp\left(-\frac{\epsilon_{j}^{\texttt{mode}}}{T}\right)\right]\,\Delta\epsilon= \int_{0}^{\infty}\,\frac{\mathcal{D}^{\texttt{mode}}(L\epsilon)}{L^{D-2}}\,\ln\left[1-\exp\left(-\frac{\epsilon}{T}\right)\right] \, d\epsilon
\end{equation}
and an asymptotic expression for $\ln Z^{\texttt{mode}}$ is obtained: 
\begin{equation}
\ln Z^{\texttt{mode}} = -L\int_{0}^{\infty}\,\mathcal{D}^{\texttt{mode}}(L\epsilon)\,\ln\left[1-\exp\left(-\frac{\epsilon}{T}\right)\right]\, d\epsilon\,,\,\,\textrm{with}\,\, L\rightarrow\infty \, .
\label{Z-general}
\end{equation}
Equation~(\ref{Z-general}) can be seen as an analogous Thomas-Fermi approximation in the AdS bosonic systems.

It is also relevant to consider not only the thermodynamic
limit, but also how this limit is approached. Hence, in the present
work, we define a quasithermodynamic limit~\cite{Maslov} as 
\begin{equation}
V_{\textrm{eff}} \,\, \textrm{large but finite, with}\,\, T\,\,\textrm{fixed}\,\,\textrm{and}\,\,\mu=0\,\,.
\end{equation}
This limit is 
particularly relevant in the present work. Unlike the usual thermodynamics in a Minkowski cavity (where the thermodynamic limit can be associated to the cavity radius going to infinity), the finiteness of the AdS radius is naturally compatible with the quasithermodynamic limit.

It is convenient to rewrite the partition function~(\ref{partion_function_ads}) in terms of the ``reduced energy'' $K_{j}^{\texttt{mode}} \equiv L\epsilon_{j}^{\texttt{mode}}$ as
\begin{equation}
\ln Z^{\texttt{mode}} = -\sum_{j}\,\mathcal{D}^{\texttt{mode}}(K_{j}^{\texttt{mode}}) \,
\ln\left[1-\exp\left(-\frac{K_{j}^{\texttt{mode}}}{LT}\right)\right] \, .
\label{Z-general-2}
\end{equation}
The elements of set $\{K_{j}^{\texttt{mode}}\}$ are integers that
do not depend on $L$ or $T$, obtained in a straightforward way from
Eq.~(\ref{general-spectra}). We conclude that the partition function $Z^{\texttt{mode}}$ depends on $L$ and $T$ as
\begin{equation}
\ln Z^{\texttt{mode}} = f(LT) \, .
\label{lnZ_f}
\end{equation}

Let us consider the high-temperature regime of the bosonic systems, that is, $LT \gg 1$. In this limit, it is useful to rewrite the integral expression for $\ln Z^{\texttt{mode}}$. Making $\epsilon=T\tilde{\epsilon}$,
\begin{equation}
L\,\int_{0}^{\infty}\,\mathcal{D^{\texttt{mode}}}(L\epsilon)\,\ln\left[1-\exp\left(-\frac{\epsilon}{T}\right)\right]\, d\epsilon
= LT\,\int_{0}^{\infty}\,\mathcal{D}^{\texttt{mode}}(LT\tilde{\epsilon})\,\ln\left[1-\exp\left(-\tilde{\epsilon}\right)\right]\, d\tilde{\epsilon}\,\,,
\end{equation}
and 
\begin{gather}
\ln Z^{\texttt{mode}} = -LT\,\int_{0}^{\infty}\,\mathcal{D}^{\texttt{mode}}(LT\tilde{\epsilon})\,\ln\left(1-e^{-\tilde{\epsilon}}\right)\, d\tilde{\epsilon}\left[1+o\left(\frac{1}{LT}\right)\right] \, .
\label{integral_formula}
\end{gather}
Strictly speaking, this integral representation for $\ln Z^{\texttt{mode}}$ is only relevant for $LT \gg 1$.
In practice, moderate values of $LT$ are enough to guarantee a reasonable applicability of the result, as discussed in subsection~\ref{numerics}.

A general expression for the partition function associated to a given
perturbation in the 
high-temperature regime, beyond the dominant
term $(LT)^{D-1}$, can be obtained. Expanding the degeneracy function~(\ref{degeneracy-function}) around $\tilde{\epsilon}=0$,
\begin{equation}
\mathcal{D}^{\texttt{mode}}\left(\tilde{\epsilon}\right) =
\sum_{i=0}^{D-2} \frac{1}{i!} \,\left. \frac{d^{i}\,\mathcal{D}^{\texttt{mode}}\left(\tilde{\epsilon}\right)}{d\tilde{\epsilon}^{i}}\right\vert _{\tilde{\epsilon}=0}\tilde{\epsilon}^{i} \, ,
\end{equation}
and using the integral representation for the Riemann zeta function
$\zeta$,
\begin{equation}
\zeta\left(s\right)=-\frac{1}{\Gamma\left(s-1\right)}\int_{0}^{\infty}t^{s-2}\ln\left(1-e^{-t}\right)\, dt\,,\ \textrm{Re}\, s>1\,,
\end{equation}
the partition function~(\ref{integral_formula}) can be written as
\begin{equation}
\ln Z^{\texttt{mode}} = \sum_{i=0}^{D-2}\left.\frac{d^{i}\,\mathcal{D}^{\texttt{mode}}\left(\tilde{\epsilon}\right)}{d\tilde{\epsilon}^{i}}\right\vert _{\tilde{\epsilon}=0}\frac{\zeta\left(i+2\right)}{i!}\Gamma\left(i+1\right)\left(LT\right)^{i+1}\,.
\end{equation}

For values of $LT$ close to $1$, the integral formula in Eq.~(\ref{integral_formula}) is not useful. In this case, expression~(\ref{Z-general-2}) for $\ln Z$ can be numerically treated. In the present work, we have conducted an extensive numerical investigation of the scalar, electromagnetic and gravitational partition functions and derived quantities. The results will be reported in the following sections.

We consider next the low-temperature limit, that is, the quasithermodynamic limit with a large effective volume and $0<LT\ll1$. In this regime, the fundamental mode of any perturbation considered dominates the sum in Eq.~(\ref{Z-general-2}). An analytic result for the partition function associated to a given mode can be determined,
\begin{gather}
\ln Z^{\texttt{mode}} = 
-\mathcal{D}^{\texttt{mode}} (K_{0}^{\texttt{mode}}) \ln\left[1-\exp\left(-\frac{K_{0}^{\texttt{mode}}}{LT}\right)\right]
= \mathcal{D}^{\texttt{mode}} (K_{0}^{\texttt{mode}})
\exp \left(-\frac{K_{0}^{\texttt{mode}}}{LT}\right) \, , \nonumber \\
0<LT\ll1 \, .
\label{lnZ_low_temperature}
\end{gather}
Explicit expressions for the scalar, electromagnetic and gravitational fields will be presented in the next subsections.

Once $Z^{\texttt{mode}}$ is obtained, it is straightforward to construct $Z^{\texttt{field}}$. And with the partition functions, all the associated thermodynamic quantities can be readily calculated. For example, entropy~($S^{\texttt{field}}$), internal energy~($U^{\texttt{field}}$) and pressure~($P^{\texttt{field}}$) are given by
\begin{gather}
S^{\texttt{field}} = \frac{\partial}{\partial T}\left(T\,\ln Z^{\texttt{field}} \right) \,,
\,\, U^{\texttt{field}} = T^{2}\frac{\partial}{\partial T}\, \left( \ln Z^{\texttt{field}} \right) \,, \nonumber \\
P^{\texttt{field}} = \frac{\partial}{\partial V_{\textrm{eff}}}\left(T\,\ln Z^{\texttt{field}} \right) 
= \frac{TL^{-(D-2)}}{D-1}\,\frac{\partial}{\partial L}\, \left( \ln Z^{\texttt{field}} \right) \, .
\label{def_P}
\end{gather}
In the next subsections, we will explore specific characteristics
of the particular bosonic fields considered.

\subsection{Thermodynamics of the massless scalar field}

\label{sec_thermo_scalar_field}

We initially consider a thermal gas of ``scalar photons'' in  anti--de
Sitter spacetime. For the massless scalar field, there is only one mode, implying that the partition function for $\texttt{field}=\textrm{sc}$ is equal to the partition function for $\texttt{mode}=\textrm{sc}$.
The degeneracy function $\mathcal{D}^{\textrm{sc}}(L\epsilon^{\textrm{sc}})$ is given by Eq.~(\ref{degenerescencia12-1}).

\begin{figure}
\begin{centering}
\includegraphics[width=0.7\textwidth]{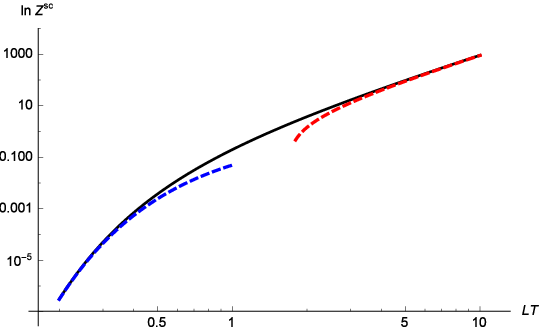} 
\end{centering}
\caption{Log-log graphs for $\ln Z^{\textrm{sc}}$ with $D=4$. The continuous line is the full numerical evaluation. The dashed lines are the low-temperature and high-temperature formulas. Results for other values of $D$ are qualitatively similar.}
\label{grafLnZsc}
\end{figure}

The low-temperature regime of the massless scalar field is characterized
by the partition function in Eq.~(\ref{lnZ_low_temperature}) with
\begin{equation}
K_{0}^{\textrm{sc}} = L \epsilon_{0}^{\textrm{sc}} = D-1 \, , \,\,
\mathcal{D}^{\textrm{sc}} \left( K_{0}^{\textrm{sc}} \right) = 1 \, .
\end{equation}
Hence, in this regime,
\begin{equation}
\ln Z^{\textrm{sc}} = \exp \left( -\frac{D-1}{LT} \right) \, ,
\end{equation}
and from this expression, 
\begin{equation}
S^{\textrm{sc}} = \frac{D-1}{LT} \, \exp\left(-\frac{D-1}{LT}\right) \, ,
\,\, U^{\textrm{sc}} = \frac{D-1}{L}\,\exp\left(-\frac{D-1}{LT}\right) \, ,
\,\,  
P^{\textrm{sc}}=\frac{1}{L^{D}}\,\exp\left(-\frac{D-1}{LT}\right) \,.
\label{ScalarLowTemperature}
\end{equation}
The low-temperature result for $\ln Z^{\textrm{sc}}$ is compared
to the numerical evaluation of the partition function in Fig.~\ref{grafLnZsc}.

The intermediate- and high-temperature regime of the scalar gas can be explored with the integral formula~(\ref{integral_formula}) for $\ln\, Z^{\textrm{sc}}$. Considering the dominant and subdominant terms (since we are assuming a finite value of $LT$ with $LT\gg1$), 
\begin{equation}
\mathcal{D}^{\textrm{sc}}(LT\tilde{\epsilon})=\frac{(LT\tilde{\epsilon})^{D-2}}{(D-2)!}\left\{ 1-\frac{\left(D-1\right)(D-2)}{2}\,\frac{1}{LT\tilde{\epsilon}}+o\left[\frac{1}{(LT\tilde{\epsilon})^{2}}\right]\right\} \,.\label{D_sc_largeT}
\end{equation}
Using Eq.~(\ref{D_sc_largeT}), the integral in Eq.~(\ref{integral_formula})
can be evaluated. The partition function in this limit is 
\begin{equation}
\ln\, Z^{\textrm{sc}} = \zeta(D)\,(LT)^{D-1} \left[ 
1 -\frac{(D-1)\zeta(D-1)}{2\zeta(D)} \,\frac{1}{LT} 
\right] \,.
\label{lnZ_sc}
\end{equation}
The high-temperature expression for $\ln Z^{\textrm{sc}}$
is compared to the numerical evaluation of the partition function in Fig.~\ref{grafLnZsc}.

It is straightforward to obtain explicit expressions for thermodynamic
quantities, once the partition function is determined. For instance,
the entropy is 
\begin{equation}
S^{\textrm{sc}} = D\zeta(D) (LT)^{D-1} \left[ 
1 - \frac{(D-1)^{2} \zeta(D-1)}{2D\,\zeta(D)} \, \frac{1}{LT}
\right] \,,
\label{entropy_ads}
\end{equation}
the internal energy is 
\begin{equation}
U^{\textrm{sc}} = (D-1) \zeta(D)T(LT)^{D-1} \left[ 
1 - \frac{(D-2)\zeta(D-1)}{2\,\zeta(D)}\,\frac{1}{LT} 
\right] \,,
\label{energyinternal_ads}
\end{equation}
and 
the pressure is
\begin{equation}
P^{\textrm{sc}} = \zeta(D) T^{D} \left[ 
1 - \frac{(D-2)\zeta(D-1)}{2\,\zeta(D)} \, \frac{1}{LT}
\right] \,.
\label{pressure_ads}
\end{equation}

\subsection{Thermodynamics of the electromagnetic field}

The next bosonic system to be considered is a thermal gas of photons.
As seen in subsection~\ref{eletromagnetic_field_sec}, the electromagnetic
perturbations can be decomposed in two independent modes, the electromagnetic
vector and scalar perturbations. Hence, the partition function 
$Z^{\textrm{el}}$ for a photon gas in $D$-dimensional AdS spacetime is given by 
\begin{equation}
\ln Z^{\textrm{el}} =
\ln Z^{\textrm{el-vc}} + \ln Z^{\textrm{el-sc}} \,,
\label{partion_function_ads_eletromag}
\end{equation}
where $Z_{D}^{\textrm{el-vc}}$ and $Z_{D}^{\textrm{el-sc}}$ are
the partition functions associated to the electromagnetic vector and
scalar modes respectively.

\begin{figure}
\begin{centering}
\includegraphics[width=0.7\textwidth]{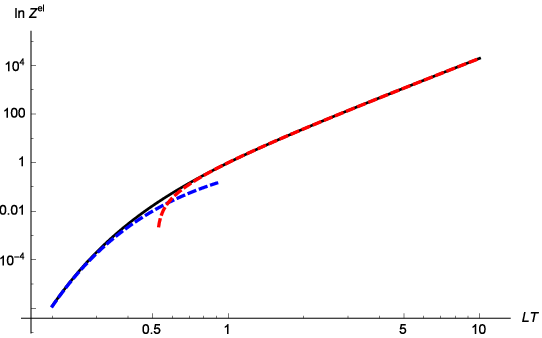} 
\end{centering}
\caption{Log-log graphs for $\ln Z^{\textrm{el}}$ with $D=5$. 
The continuous line is the full numerical evaluation. The dashed lines are the low-temperature and high-temperature formulas. Results for other values of $D$ are qualitatively similar.}
\label{grafLnZ-el} 
\end{figure}

In the low-temperature regime, the values for the constants 
$K_{0}^{\texttt{mode}} = L \epsilon_{0}^{\texttt{mode}}$
associated to the electromagnetic field are 
\begin{equation}
K_{0}^{\textrm{el-vc}} = D-1 \, , \,\,
K_{0}^{\textrm{el-sc}} = \frac{D+1+|D-5|}{2} \, , \,\,
\mathcal{D}^{\textrm{el-vc}} \left( K_{0}^{\textrm{el-vc}} \right) = 
\mathcal{D}^{\textrm{el-sc}} \left( K_{0}^{\textrm{el-sc}} \right) =
D-1 \, .
\end{equation}
In the four-dimensional case, $K_{0}^{\textrm{el-vc}} = K_{0}^{\textrm{el-sc}}$ and both scalar and vector modes contribute. In higher dimensions, $K_{0}^{\textrm{el-sc}}<K_{0}^{\textrm{el-vc}}$ and the scalar mode dominates. It follows from Eq.~(\ref{lnZ_low_temperature}) that the leading contribution for the electromagnetic partition function in the limit $T\rightarrow0$ is
\begin{equation}
\ln Z^{\textrm{el}}=\begin{cases}
6 \exp\left(-\frac{3}{LT}\right) & \, , \, D=4 \\
(D-1) \exp\left(- \frac{D-2}{LT} \right) & \, , \, D>4
\end{cases} \,.
\label{lnZ-el-lowT}
\end{equation}
The low-temperature analytic expression for $\ln Z^{\textrm{el}}$ is compared to the numerical evaluation of the partition function in Fig.~\ref{grafLnZ-el}.
From expression~(\ref{lnZ-el-lowT}), some thermodynamic quantities
for the AdS photon gas in the low-temperature regime are
\begin{equation}
S^{\textrm{el}}=\begin{cases}
\frac{18}{LT} \, \exp\left(-\frac{3}{LT}\right) & \, , \, D=4\\
\frac{(D-1)(D-2)}{LT} \, \exp\left(-\frac{D-2}{LT}\right) & \, , \, D>4
\end{cases}\,,
\label{ElectromagneticLowTemperature}
\end{equation}
\begin{equation}
U^{\textrm{el}}=\begin{cases}
\frac{18}{L}\,\exp\left(-\frac{2}{LT}\right) & \, , \, D=4\\
\frac{(D-1)(D-2)}{L}\,\exp\left(-\frac{D-2}{LT}\right) & \, , \, D>4
\end{cases}\,,
\end{equation}
\begin{equation}
P^{\textrm{el}}=\begin{cases}
\frac{6}{L^{4}}\,\exp\left(-\frac{3}{LT}\right) & \, , \, D=4\\
\frac{D-2}{L^{D}}\,\exp\left(-\frac{D-2}{LT}\right) & \, , \, D>4
\end{cases}\,.
\end{equation}

In the high-temperature regime, the degeneracies of the vector and
scalar electromagnetic modes can be written as 
\begin{equation}
\mathcal{D}^{\textrm{el-vc}}(LT\tilde{\epsilon})=\frac{(LT\tilde{\epsilon})^{D-2}}{(D-2)!}\left\{ 1-\frac{\left(D-2\right)(D-3)}{2}\,\frac{1}{LT\tilde{\epsilon}}+o\left[\frac{1}{(LT\tilde{\epsilon})^{2}}\right]\right\} \,,
\end{equation}
\begin{equation}
\mathcal{D}^{\textrm{el-sc}}(LT\tilde{\epsilon})=\begin{cases}
\frac{(LT\tilde{\epsilon})^{2}}{2}\left\{ 1-\frac{1}{LT\tilde{\epsilon}}+o\left[\frac{1}{(LT\tilde{\epsilon})^{2}}\right]\right\}  & ,\,\, D=4\\
\frac{(LT\tilde{\epsilon})^{3}}{6}\left\{ 1-\frac{1}{(LT\tilde{\epsilon})^{2}}+o\left[\frac{1}{(LT\tilde{\epsilon})^{3}}\right]\right\}  & ,\,\, D=5\\
\frac{(LT\tilde{\epsilon})^{D-2}}{(D-2)!}\left\{ 1-\frac{\left(D-2\right)(D-5)}{2}\,\frac{1}{LT\tilde{\epsilon}}+o\left[\frac{1}{(LT\tilde{\epsilon})^{2}}\right]\right\} \, & ,\,\, D>5
\label{deg-el-sc-highT}
\end{cases}\,.
\end{equation}
It should be noticed that in the $D=5$ case the subdominant term in Eq.~(\ref{deg-el-sc-highT}) is of the order of $1/(LT)^{2}$.
The calculation of the associated partition functions is straightforward:
\begin{equation}
\ln\, Z^{\textrm{el-vc}} =  \zeta(D) \, (LT)^{D-1} \left[ 
1 - \frac{(D-3)\zeta(D-1)}{2\zeta(D)} \, \frac{1}{LT}
\right] \, ,
\label{lnZ-el-vc-highT}
\end{equation}
\begin{equation}
\ln\, Z^{\textrm{el-sc}} = \begin{cases}
\zeta(4)\,(LT)^{3} \left[ 
1 - \frac{\zeta(3)}{2\zeta(4)}\,\frac{1}{LT}
\right]  & ,\,\, D=4\\
\zeta(5)\,(LT)^{4} \left[ 
1 - \frac{\zeta(3)}{6\zeta(5)}\,\frac{1}{(LT)^{2}}
\right]  & ,\,\, D=5\\
\zeta(D)\,(LT)^{D-1} \left[ 
1 - \frac{(D-5)\zeta(D-1)}{2\zeta(D)}\,\frac{1}{LT}
\right]  & ,\,\, D>5
\end{cases}\,.
\label{lnZ-el-sc-highT}
\end{equation}
With expressions~(\ref{lnZ-el-vc-highT}) and (\ref{lnZ-el-sc-highT}),
we obtain the partition function of the photon gas in the high-temperature
regime: 
\begin{equation}
\ln\, Z^{\textrm{el}}=\begin{cases}
2\zeta(4)\,(LT)^{3} \left[ 
1 - \frac{\zeta(3)}{2\zeta(4)}\,\frac{1}{LT} 
\right]  & ,\,\, D=4\\
2\zeta(D)\,(LT)^{D-1} \left[ 
1 - \frac{(D-4)\zeta(D-1)}{2\zeta(D)}\,\frac{1}{LT}
\right]  & ,\,\, D>4
\end{cases}\,.\label{lnZ-el-highT}
\end{equation}
The expression for $\ln\, Z^{\textrm{el}}$ in Eq.~(\ref{lnZ-el-highT})
is illustrated in Fig.~\ref{grafLnZ-el}, where the analytic expression is compared to the numerical evaluation.

With the formula~(\ref{lnZ-el-highT}), the entropy, internal
energy and pressure of the photon gas can be derived:
\begin{equation}
S^{\textrm{el}} = \begin{cases}
8\zeta(4)(LT)^{3} \left[ 
1-\frac{3\zeta(3)}{8\zeta(4)}\,\frac{1}{LT}
\right]  & ,\,\, D=4 \\
2D\zeta(D)(LT)^{D-1} \left[ 
1-\frac{(D-1)(D-4)\zeta(D-1)}{2D\zeta(D)}\,\frac{1}{LT}
\right]  & ,\,\, D>4
\end{cases} 
\, ,
\end{equation}
\begin{equation}
U^{\textrm{el}} = \begin{cases}
6\zeta(4)T(LT)^{3} \left[ 
1-\frac{\zeta(3)}{3\zeta(4)}\,\frac{1}{LT}
\right]  & ,\,\, D=4\\
2(D-1)\zeta(D)T(LT)^{D-1} \left[ 
1-\frac{(D-2)(D-4)\zeta(D-1)}{2(D-1)\zeta(D)}\,\frac{1}{LT}
\right]  & ,\,\, D>4
\end{cases} \,,
\end{equation}
\begin{equation}
P^{\textrm{el}} = \begin{cases}
2\zeta(4) T^{4} \left[ 
1-\frac{\zeta(3)}{3\zeta(4)}\,\frac{1}{LT}
\right]  & ,\,\, D=4\\
2\zeta(D) T^{D} \left[ 
1-\frac{(D-2)(D-4)\zeta(D-1)}{2(D-1)\zeta(D)}\,\frac{1}{LT}
\right]  & ,\,\, D>4
\end{cases}\,.
\end{equation}

\subsection{Thermodynamics of the gravitational field}

It is finally considered a thermal graviton gas in $D$-dimensional
AdS spacetime. As seen in subsection~\ref{gravitational_perturbations_sec},
gravitational perturbations can be decoupled in three independent
modes (tensor, vector and scalar). In the four-dimensional case, the
gravitational tensor mode does not contribute, and the partition function
for the graviton gas is expressed as 
\begin{equation}
\ln Z^{\textrm{gr}} =
\ln Z^{\textrm{gr-vc}}+\ln Z^{\textrm{gr-sc}}\,,\,\,\textrm{for}\, D=4 \, .
\label{partion_function_ads_grav-1}
\end{equation}
In five or higher dimensions, all three gravitational modes are relevant,
and the gravitational field partition function is written as 
\begin{equation}
\ln Z^{\textrm{gr}} =
\ln Z^{\textrm{gr-t}} + \ln Z^{\textrm{gr-vc}} + \ln Z^{\textrm{gr-sc}}
\,,\,\,\textrm{for} \, D>4\, .
\label{partion_function_ads_grav-2}
\end{equation}

Let us consider the low-temperature limit. Carrying out
a development very similar to the one presented in the previous subsections, we observe that
\begin{gather}
K_{0}^{\textrm{el-sc}} = \frac{D+3+|D-5|}{2} \, , \,\,
K_{0}^{\textrm{gr-vc}} = D \, , \,\,
K_{0}^{\textrm{gr-t}} = D+1 \, , \,\, 
\nonumber \\
\mathcal{D}^{\textrm{el-sc}} \left( K_{0}^{\textrm{el-sc}} \right) = 
\mathcal{D}^{\textrm{gr-vc}} \left( K_{0}^{\textrm{gr-vc}} \right) =
\mathcal{D}^{\textrm{gr-t}} \left( K_{0}^{\textrm{gr-t}} \right) =
\frac{D(D-1)}{2}
\,.
\end{gather}
If $D=4$, both scalar and vector modes are relevant in the low-temperature regime. For higher dimensions, the scalar mode dominates.

\begin{figure}
\begin{centering}
\includegraphics[width=0.7\textwidth]{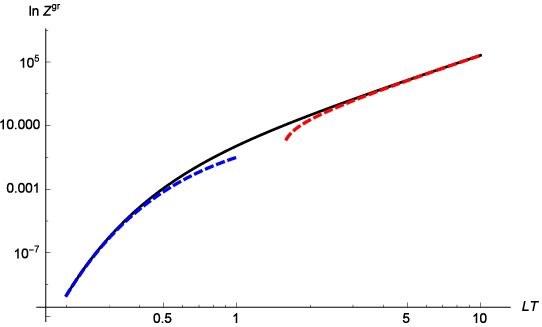} 
\end{centering}
\caption{Log-log graphs for $\ln Z^{\textrm{gr}}$ with $D=6$. 
The continuous line is the full numerical evaluation. The dashed lines are the low-temperature and high-temperature formulas. Results for other values of $D$ are qualitatively similar.}
\label{grafLnZ-gr} 
\end{figure}

Using Eq.~(\ref{lnZ_low_temperature}), we obtain in the limit $T \rightarrow 0$ that
\begin{equation}
\ln Z^{\textrm{gr}}=\begin{cases}
12\exp\left(-\frac{4}{LT}\right) & \,, \,\, D=4\\
\frac{D(D-1)}{2} \exp\left(-\frac{D-1}{LT}\right) & \,, \,\, D>4
\end{cases} \, .
\label{lnZ-gr-lowtemperature}
\end{equation}
Result~(\ref{lnZ-gr-lowtemperature}) for $\ln Z^{\textrm{gr}}$ in the low-temperature limit is illustrated in Fig.~\ref{grafLnZ-gr}, where the analytic result is compared to the numerical development. Entropy, internal energy and pressure are readily calculated: 
\begin{equation}
S^{\textrm{gr}}=\begin{cases}
\frac{48}{LT} \, \exp\left(-\frac{4}{LT}\right) & \,, \,\, D=4\\
\frac{D(D-1)^{2}}{2LT} \, \exp\left(-\frac{D-1}{LT}\right) & \,, \,\, D>4
\end{cases} \, ,
\label{GravitationalLowTemperature}
\end{equation}
\begin{equation}
U^{\textrm{gr}}=\begin{cases}
\frac{48}{L} \, \exp\left(-\frac{4}{LT}\right) & \,, \,\, D=4\\
\frac{D(D-1)^{2}}{2L} \, \exp\left(-\frac{D-1}{LT}\right) & \,, \,\, D>4
\end{cases}\,,
\end{equation}
\begin{equation}
P^{\textrm{gr}}=\begin{cases}
\frac{16}{L^{4}} \, \exp\left(-\frac{4}{LT}\right) & \,, \,\, D=4\\
\frac{D(D-1)}{2L^{D}} \, \exp\left(-\frac{D-1}{LT}\right) & \,, \,\, D>4
\end{cases}\,.
\end{equation}

For the intermediate- and high-temperature regimes, considering that
$\mathcal{D}^{\textrm{gr-t}}(LT\tilde{\epsilon})=\mathcal{D}^{\textrm{sc}}(LT\tilde{\epsilon})$,
$\mathcal{D}^{\textrm{gr-vc}}(LT\tilde{\epsilon})=\mathcal{D}^{\textrm{el-vc}}(LT\tilde{\epsilon})$
and $\mathcal{D}^{\textrm{gr-sc}}(LT\tilde{\epsilon})=\mathcal{D}^{\textrm{el-sc}}(LT\tilde{\epsilon})$,
combined with results from previous subsections, the calculation of
the associated partition functions is straightforward:
\begin{equation}
\ln\, Z^{\textrm{gr}}=\begin{cases}
2\zeta(4)\,(LT)^{3}\left[ 
1-\frac{\zeta(3)}{2\zeta(4)}\,\frac{1}{LT}
\right] \, & ,\,\, D=4\\
3\zeta(D)(LT)^{D-1}\left[ 
1-\frac{(D-3)\zeta(D-1)}{2\zeta(D)}\frac{1}{LT}
\right] \, & ,\,\, D>4
\end{cases} \, .
\label{lnZ-gr-highT}
\end{equation}
Expression~(\ref{lnZ-gr-highT}) for $\ln\, Z^{\textrm{gr}}$ is illustrated in Fig.~\ref{grafLnZ-gr}, where the analytic result is compared to the numerical calculation. From Eq.~(\ref{lnZ-gr-highT}), the entropy, internal energy
and pressure associated to the AdS graviton gas are calculated:
\begin{equation}
S^{\textrm{gr}}=\begin{cases}
8\zeta(4)(LT)^{3} \left[
1-\frac{3\zeta(3)}{8\zeta(4)}\,\frac{1}{LT}
\right]  & ,\,\, D=4\\
3D\zeta(D)(LT)^{D-1} \left[ 
1 - \frac{(D-3)(D-1)\zeta(D-1)}{2D\zeta(D)}\frac{1}{LT}
\right]  & ,\,\, D>4
\end{cases} \, ,
\end{equation}
\begin{equation}
U^{\textrm{gr}}=\begin{cases}
6\zeta(4)T(LT)^{3} \left[ 
1 - \frac{\zeta(3)}{3\zeta(4)}\,\frac{1}{LT}
\right]  & ,\,\, D=4\\
3(D-1)\zeta(D)\, T(LT)^{D-1} \left[ 
1 - \frac{(D-3)(D-2)\zeta(D-1)}{2(D-1)\zeta(D)}\frac{1}{LT}
\right]  & ,\,\, D>4
\end{cases} \, ,
\end{equation}
\begin{equation}
P^{\textrm{gr}}=\begin{cases}
2\zeta(4) T^{4} \left[ 
1-\frac{\zeta(3)}{3\zeta(4)}\,\frac{1}{LT}
\right]  & ,\,\, D=4\\
3\zeta(D) T^{D} \left[ 
1-\frac{(D-3)(D-2)\zeta(D-1)}{2(D-1)\zeta(D)}\frac{1}{LT}
\right]  & ,\,\, D>4
\end{cases}\, .
\end{equation}

\section{Analysis of the thermodynamics}
\label{analysis}

In this section, we will explore general characteristics of the thermodynamics associated to bosonic gases in anti--de Sitter spacetime. Structural aspects, numerical results, instabilities, and the issue of a proper thermodynamic volume will be explored.

\subsection{Structural aspects}

%
Quasihomogeneity, that is, homogeneity of degrees different of one and zero, is an essential characteristic of a well-defined equilibrium thermodynamics.
While quasihomogeneity is explicit in the strict thermodynamic limit, it is not so beyond this limit, in the quasithermodynamic regime. However, the partition functions can be written as $Z^{\texttt{field}} = f(LT)$, as seen in Eq.~(\ref{lnZ_f}).
It follows that quasihomogeneity is recovered, demanding that $LT$ should be a homogeneous function of degree zero (otherwise, different powers of $LT$ will give different degrees). We note that it is the same requirement due to scaling arguments for the consistency of the Schwarzschild-anti de Sitter thermodynamics~\cite{Baldiotti:2016lcf,Baldiotti:2017ywq,kastor2010}.

The condition $Z^{\texttt{field}} = f(LT)$, besides ensuring quasihomogeneity, also furnishes a consistency result. This restriction on the form of the function $Z^{\texttt{field}}$ implies that 
\begin{equation}
P^{\texttt{field}} =
\frac{T^{2}}{(D-1)\, L^{D-2}}\,\frac{d}{d(LT)} f(LT)
= \frac{U^{\texttt{field}}}{(D-1)\, L^{D-1}}\,,
\end{equation}
where the expressions in~(\ref{def_P}) were used. Taking
into account our definition of $V_{\textrm{\textrm{eff}}}$ in Eq.~(\ref{volumeeffn}),
\begin{equation}
P^{\texttt{field}} =
\frac{1}{D-1}\,\frac{U^{\texttt{field}}}{V_{\textrm{\textrm{eff}}}} \,.
\label{state_eq_null_fluid}
\end{equation}
The equation of state~(\ref{state_eq_null_fluid}) shows that the
massless bosonic gases behave as a null fluid, as expected. We conclude
that the proposal for the effective volume in Eq.~(\ref{volumeeffn})
is compatible with the general thermodynamic description for the bosonic
systems treated in the present work. More comments on this topic will be presented in subsection~\ref{volume-homogeneity}.

Let us consider the low-temperature limit of a massless bosonic field
in AdS spacetime. Taking into account
results~(\ref{ScalarLowTemperature}), (\ref{ElectromagneticLowTemperature}) and (\ref{GravitationalLowTemperature}), it follows that the entropy behaves as
\begin{equation}
S^{\texttt{field}} \propto \, 
\exp \left(-\frac{\mathcal{K}_{S}^{\texttt{field}}}{LT}\right)
\,\, , \,\, 0<LT\ll1 \, ,
\label{S_low_temperature}
\end{equation}
where $\mathcal{K}_{S}^{\texttt{field}}$ is a positive constant which depends on the particular field and dimension considered. Entropy tends to zero as the temperature approaches absolute zero, and therefore the AdS bosonic systems respect the third law of thermodynamics. 

In the high-temperature regime, the general qualitative behavior of the AdS bosonic systems can be analyzed.
With the integral formula~(\ref{integral_formula}) for the partition functions
and 
$\mathcal{D}^{\texttt{mode}}(LT\,\tilde{\epsilon}) =
(LT\,\tilde{\epsilon})^{D-2} \left[ 1+o\left(\frac{1}{LT}\right)\right] $
for large (but finite) $LT$, we obtain 
\begin{equation}
\ln Z^{\texttt{field}} =
-(LT)^{D-1} \, \int_{0}^{\infty} \, \ln\left[1-\exp\left(-\tilde{\epsilon}\right)\right]\, d\tilde{\epsilon}\propto T^{D-1}\, V_{\textrm{eff}}\,.
\label{lnZ_largeT}
\end{equation}
Therefore,
\begin{equation}
U^{\texttt{field}} \propto 
P^{\texttt{field}} \propto 
T^{D}\,.
\label{U_P_largeT}
\end{equation}
Result~(\ref{U_P_largeT}) indicates that the bosonic fields in AdS spacetime obey the $D$-dimensional Stefan-Bolzmann law in the thermodynamic limit.

It is instructive to consider the energy of a given mode radiated per unit frequency, the AdS-analogous Planck formula. In the quasithermodynamic limit, using Eqs.~(\ref{integral_formula}) and~(\ref{def_P}),
\begin{equation}
U^{\mathtt{\operatorname{mod}e}}\left(  T\right) = T^{2}\left\{  -L
\int_{0}^{\infty}\,\mathcal{D}^{\mathtt{mode}}(L\epsilon)\,\frac{\partial
}{\partial T}\ln\left[  1 - \exp \left(- \frac{\epsilon}{T} \right) \right]  \,d\epsilon\right\}
= \int_{0}^{\infty}\,\mathcal{D}^{\mathtt{mode}}(L\epsilon)\frac{L\epsilon
}{\exp \left(\frac{\epsilon}{T} \right) - 1}\,d\epsilon \, .
\end{equation}
It follows that the spectral distribution of energy $\tilde{U}^{\text{mode}}(T,\epsilon)$, with the angular frequency $\omega = \epsilon$, is given by
\begin{equation}
\tilde{U}^{\text{mode}}(T,\epsilon) =
\frac{\mathcal{D}^{\mathtt{mode}}(L\epsilon)\, L\epsilon}{\exp \left(\frac{\epsilon}{T} \right)-1}
\, .
\end{equation}
We define the energy density per unit frequency as
\begin{equation}
\tilde{u}^{\text{mode}}(T,\epsilon) \equiv
\frac{\tilde{U}^{\text{mode}}(T,\epsilon)}{V_{\textrm{eff}}} \, .
\end{equation}
The quantity $\tilde{u}^{\text{mode}}(T,\epsilon)$ will furnish the AdS-analogous Planck formulas.
For example, the explicit expression for $\tilde{u}^{\textrm{sc}}(T,\epsilon)$ is
\begin{equation}
\tilde{u}^{\textrm{sc}} (T,\epsilon) =
\frac{1}{\left(  D-2\right)!} \left[  1- \frac{(D-1)  (D - 2)}{2}
\frac{1}{L\epsilon}\right]  
\frac{\epsilon^{D-1}}{\exp \left(\frac{\epsilon}{T} \right) - 1} \, .
\label{u-scalar}
\end{equation}
From Eq.~(\ref{u-scalar}), the relative difference between the spectral energy density, considering the AdS and the usual cavity thermodynamics, is of the order of $1/(L\epsilon)$. Similar results can be obtained for the other modes and fields.

\subsection{Numerical results}
\label{numerics}

We performed an extensive numerical investigation of the partition functions and associated quantities presented in previous section. 
In addition to the field label $\{ \textrm{sc},\textrm{el},\textrm{gr} \}$, dimension $D$, AdS radius $L$ and temperature $T$, the numerical analysis introduces new parameters for the actual (numerical) calculations: the maximum number of energy levels $N_{\textrm{max}}$ considered in the sum~(\ref{partion_function_ads}) and the approximate number of digits $N_{\textrm{prec}}$ in floating-point operations. Alternatively, instead of $N_{\textrm{max}}$, one could use the maximum values for the multipole and overtone numbers, denoted by $\ell_{\textrm{max}}$ and $n_{\textrm{max}}$, respectively. 
We considered scalar, electromagnetic and gravitational fields with spacetime dimension varying from 4 to 10. 
The ranges of $LT$ used in this work were selected in order to include both the low- and high-temperature regimes. In these limits, the numerical and analytic results were compared. Usual choices of $LT$ ranges included the interval $(0.1,15)$.
Typical calculations were performed with $\ell_{\textrm{max}}=50$ and $n_{\textrm{max}}=50$. Attempting to minimize the rounding error, we used%
\footnote{The numerical calculations were performed within the \textit{Mathematica} environment, where the parameter $N_{\textrm{prec}}$ can be arbitrarily fixed. However, the computational effort can be significant with a large value of $N_{\textrm{prec}}$.}
a high precision scheme with $N_{\textrm{prec}}=50$.

\begin{figure}
\begin{centering}
\includegraphics[width=0.7\textwidth]{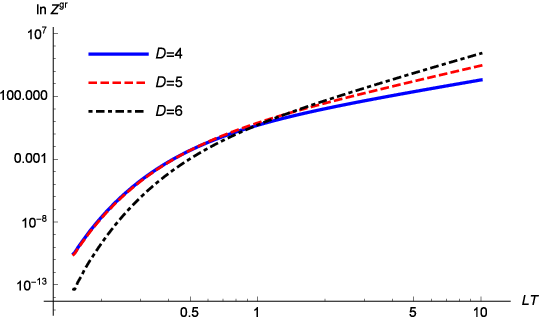} 
\end{centering}
\caption{Log-log graphs for $\ln Z^{\textrm{gr}}$ as a function of $LT$
with $D=4,5,6$. Results for other fields and dimensions are qualitatively similar.}
\label{graf-lnZ-gr-D456} 
\end{figure}

We observed that the functions $\ln Z^{\texttt{field}}$ are monotonically increasing in the dimensionless parameter $LT$, growing exponentially with low $LT$ and as a power law with high $LT$. The intermediate-temperature regime smoothly connects the low- and high-temperature limits. No discontinuities or divergences were observed in these quantities. This result is illustrated in Fig.~\ref{graf-lnZ-gr-D456}, where graphs for $\ln Z^{\texttt{gr}}$ are presented for several values of $D$. The qualitative behavior is the same for other fields and dimensions. As expected, the analytic expressions in the low- and high-temperature limits are compatible with the numerical results, as illustrated in Figs.~\ref{grafLnZsc}, \ref{grafLnZ-el} and \ref{grafLnZ-gr}.

An issue is the convergence of the numerical and analytic results produced. Specifically, we focus on the partition functions written as discrete sums, and the integral versions of those quantities obtained by analogous Thomas-Fermi approximations. The integral formulas for $\ln Z^{\texttt{field}}$ are good
approximations for the partition functions if $LT$ is higher than a given number (an analogous Thomas-Fermi critical temperature). Typically, good concordance is already observed with $LT \approx 5$.

\begin{figure}
\begin{centering}
\includegraphics[width=0.7\textwidth]{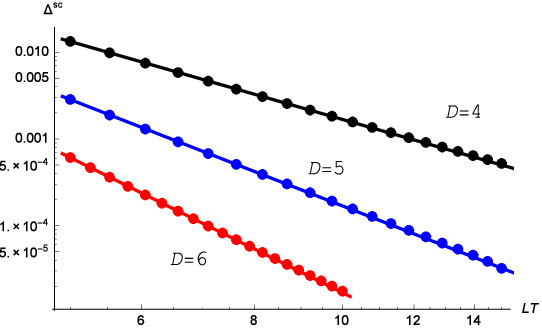} 
\end{centering}
\caption{Log-log graphs for $\Delta^{\textrm{sc}}$ in terms of $LT$, considering the scalar gas with $D=4,5,6$. The bullets are the numerical results, and
the straight lines are the power-law fits. For $D=4,5,6$, the fits are $\Delta^{\textrm{sc}} = 1.52\times(LT)^{-2.95}$,
$\Delta^{\textrm{sc}} = 1.78\times(LT)^{-4.01}$ and $\Delta^{\textrm{sc}} = 2.08\times(LT)^{-5.08}$, respectively.}
\label{grafconverge} 
\end{figure}

More precisely,  convergence can be characterized by the behavior of
the relative difference $\Delta^{\texttt{mode}}$ between the expressions defined with sums and integrals: 
\begin{equation}
\Delta^{\texttt{mode}} \equiv \left| \frac{\ln Z^{\texttt{mode}}_{\textrm{sum}} - \ln Z^{\texttt{mode}}_{\textrm{integral}}}{\ln Z^{\texttt{mode}}_{\textrm{sum}}} \right| \, .
\label{Delta}
\end{equation}
Our numerical results indicate that $\Delta^{\texttt{mode}}$ tends to zero as a power law in the form 
\begin{equation}
\Delta^{ \texttt{mode} } \propto \frac{1}{(LT)^{D-1} } \, .
\end{equation}
We illustrate the dependence of $\Delta^{ \texttt{mode} }$ with $LT$ in Fig.~\ref{grafconverge}, considering the scalar gas. Results for the other modes and fields are qualitatively similar.

\begin{figure}
\begin{centering}
\includegraphics[width=0.7\textwidth]{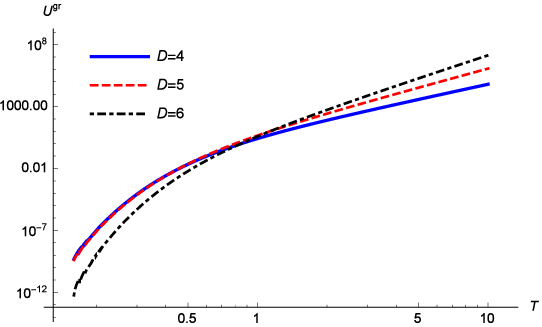} 
\end{centering}
\caption{Log-log graphs for $U^{\textrm{gr}}$ as a function of the temperature
with $D=4,5,6$. In the graphs, $L=1$. The results are qualitatively similar for other fields and values of $D$ and $L$.}
\label{graf-U-gr-D456} 
\end{figure}

\begin{figure}
\begin{centering}
\includegraphics[width=0.7\textwidth]{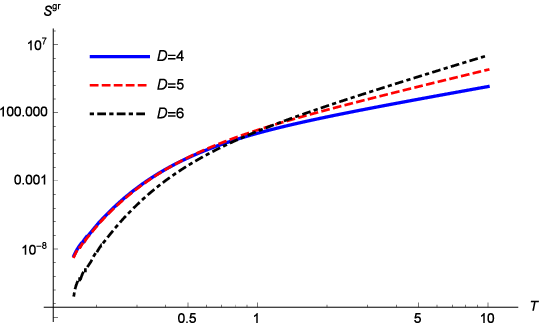} 
\end{centering}
\caption{Log-log graphs for $S^{\textrm{gr}}$ as a function of the temperature
with $D=4,5,6$. In the graphs, $L=1$. The results are qualitatively similar for other fields and values of $D$ and $L$.}
\label{graf-S-gr-D456} 
\end{figure}

\begin{figure}
\begin{centering}
\includegraphics[width=0.7\textwidth]{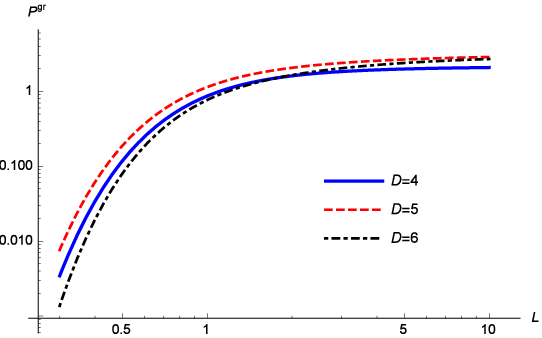} 
\end{centering}
\caption{Log-Log graphs for $P^{\textrm{gr}}$ as a function of $L$
with $D=4,5,6$. In the graphs, $T=1$. The results are qualitatively similar for other fields and values of $D$ and $L$.}
\label{graf-P-gr-D456} 
\end{figure}

Concerning the internal energy, entropy and pressure, the general behavior of these quantities is largely independent of the field and dimension.
The functions $U^{\texttt{field}}$, $S^{\texttt{field}}$ and $P^{\texttt{field}}$ are positive definite and tend to zero as $T$ tend to zero. With a constant effective volume,  $U^{\texttt{field}}$ and $S^{\texttt{field}}$ are monotonically increasing in $T$, growing as a near exponential with low $T$ and as a power law with high $T$. With the temperature constant, $P^{\texttt{field}}$ grows monotonically as the effective volume increases. 
These characteristics are illustrated in Figs.~\ref{graf-U-gr-D456}-\ref{graf-P-gr-D456}.

\subsection{Thermodynamic instabilities}

\begin{figure}
\begin{centering}
\includegraphics[width=0.7\textwidth]{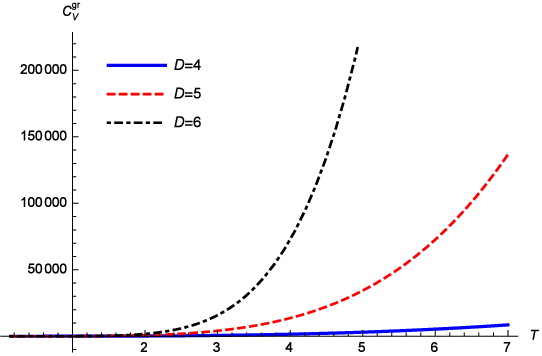} 
\end{centering}
\caption{Graphs for $C_{V}^{\textrm{gr}}$ as a function of the temperature
with $D=4,5,6$. In the graphs, $L=1$. The qualitative behavior is the same for other fields and dimensions.}
\label{graf-C-gr-D456} 
\end{figure}

Thermodynamic instabilities are identified by the behavior of quantities such as the thermal capacities and compressibility. 
Let us initially consider the thermal capacity at constant volume ($C_{V}^{\texttt{field}}$), defined as
\begin{equation}
C_{V}^{\texttt{field}} \equiv
\left. \frac{\partial U^{\texttt{field}}}{\partial T} \right|_{V_{\text{eff}}} \, .
\label{Cv}
\end{equation}
Since the functions $U^{\texttt{field}}$ with constant $L$ are monotonically increasing, $C_{V}^{\texttt{field}}$ is always positive. Therefore, the system is stable for any constant volume process. Sample graphs for the thermal capacities are presented in Fig.~\ref{graf-C-gr-D456}.

\begin{figure}
\begin{centering}
\includegraphics[width=0.7\textwidth]{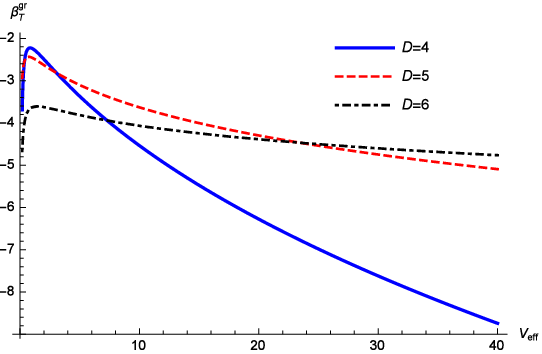} 
\end{centering}
\caption{Graphs for $\beta_{T}^{\textrm{gr}}$ as a function of the $V_{\textrm{eff}}$
with $D=4,5,6$. In the graphs, $T=1$. The qualitative behavior is the same for other fields and dimensions.}
\label{graf-beta-gr-D456} 
\end{figure}

We consider now the isothermal compressibility $\beta_{T}^{\texttt{field}}$, defined as
\begin{equation}
\beta_{T}^{\texttt{field}} \equiv
-\frac{1}{V_{\text{eff}}} 
\left.  \frac{\partial V_{\text{eff}}}{\partial P^{\texttt{field}}} \right \vert _{T}
\, .
\label{compres}
\end{equation}
Our numerical results indicate that the pressure $P^{\texttt{field}}$ is a monotonically increasing function in $V_{\text{eff}}$ (with constant $T$). It follows that $\beta_{T}^{\texttt{field}}$ is always negative. This means that the system is unstable under isothermic processes (that is, considering only variations in the cosmological constant). The general behavior of $\beta_{T}^{\texttt{field}}$ is presented in Fig.~\ref{graf-beta-gr-D456}.

The instability generated by $\beta_{T}^{\texttt{field}}<0$ is expected, since the usual photon gas in a cavity also presents this phenomenon \cite{Lef2015}.
For instance, let us consider the high-temperature regime. In the thermodynamic limit, the isothermal compressibility can be written as
\begin{equation}
\beta_{T}^{\texttt{field}} = - \mathcal{K}_{\beta}^{\texttt{field}}  \frac{LT}{T^{D}} \, ,
\label{beta_highT}
\end{equation}
where $\mathcal{K}_{\beta}^{\texttt{field}}$ is a positive constant which depends on the dimensionality and of the particular bosonic perturbation. Expression~(\ref{beta_highT}) indicates that there is no contribution for the compressibility from the dominant term. 
That is, the subdominant term dictates the behavior of $\beta_{T}^{\texttt{field}}$ in the high-temperature limit.
A negative subdominant term implies that the pressure increases with the volume. This is the observed (and perhaps counterintuitive) behavior in the AdS bosonic gases. Considering the usual photon gas, published results (for example \cite{Lef2015,Baltes1973}) indicate that the subdominant term is negative, and the behavior of the usual gas coincides with the behavior of the AdS gas as described here.%
\footnote{We remark that there is some controversy in the literature concerning the sign of this subdominant term in the usual Minkowskian setup (see for example Ref.~\cite{Gusev}).}

We can compare the corrections in the Stefan-Boltzmann law in AdS spacetime with the photon gas confined in a cavity (considering Minkowski spacetime). In the high-temperature regime, the energy density of the several bosonic systems considered have the form:
\begin{equation}
\frac{U^{\texttt{field}}}{V_{\textrm{eff}}}
= 
a^{\texttt{field}}_{D} T^{D} \left(  1-\frac{b^{\texttt{field}}_{D}}{T}
V_{\textrm{eff}}^{-\frac{1}{D-1}}\right) \, .
\end{equation}
Explicit expressions for the constants $a^{\texttt{field}}_{D}$ and $b^{\texttt{field}}_{D}$ can be readily determined. For example, considering the electromagnetic field in four dimensions,
\begin{equation}
\frac{U^{\texttt{field}}}{V_{\textrm{eff}}} =
\frac{\pi^{4}}{15}\, T^{4} - 2\zeta(3)  V_{\textrm{eff}}^{-\frac{1}{3}} \, T^{3} \,.
\label{mcu}
\end{equation}
Expression~(\ref{mcu}) can be compared to the analogous result concerning the photon gas in a cavity, which has the form \cite{BalHi76}
\begin{equation}
\frac{U^{\texttt{field}}}{V_{\textrm{eff}}} = \lambda_{1} T^{4} - \lambda_{2} T^{2} \, .
\label{usual}
\end{equation}
We observe that in the usual ``Minkowskian cavity'' case, Eq.~(\ref{usual}), the term proportional to $T^{3}$ is not present, indicating that there is no contribution proportional to the cavity area in the energy distribution \cite{CasCh70}. The conclusions from this particular case can be generalized for other fields and dimensions. The presented results show that the AdS bosonic systems 
have different behavior than the usual photon gas
in a cavity, even in the thermodynamic limit.

\subsection{Effective volume and homogeneity}
\label{volume-homogeneity}

In Sec.~\ref{background}, the effective volume in AdS spacetime was introduced within a geometric context. The goal here is
to show that this choice of thermodynamic volume is a consequence of the
homogeneity of the equations of states in the thermodynamic limit.

Let us consider (in the internal energy description) that the
bosonic systems in AdS background have two independent thermodynamic
variables: the entropy $S$ and a still not determined thermodynamic volume $V$. 
Besides, as usual, in the free energy $F=-T\ln Z=U-TS$, the independent variables are $T$ and $V$. We want to show that $V$ is given by the effective volume $V_{\text{eff}} \equiv L^{D-1}$ introduced in Sec.~\ref{background}. 
We are assuming that, in the thermodynamic limit, entropy and volume are
extensive variables (first order homogeneous), while temperature and
pressure are intensive variables (zeroth order homogeneous). 

Considering the hypothesis made, the Euler theorem for homogeneous functions applies, giving 
\begin{equation}
U^{\text{field}} = TS^{\text{field}} - PV \, .  
\label{euler}
\end{equation}
In previous expression, $P$ is the conjugate variable (to be determined)
associated to the volume $V$. Combining Eq.~(\ref{euler}) with relations~(\ref{def_P}), we obtain
\begin{equation}
PV = T \ln Z^{\text{field}} \,.
\end{equation}

From Sec.~\ref{sec_thermodynamics}, the partition function of a given
bosonic field in the thermodynamic limit can be written as 
\begin{equation}
\ln Z^{\text{field}} = \mathcal{A} \, (LT)^{D-1} \,,  
\label{1}
\end{equation}
where $\mathcal{A}$ is a constant which depends on the
particular field considered. Using expression~(\ref{1}), 
\begin{equation}
P =  \frac{\mathcal{A} T}{V}(LT)^{D-1} \,.  
\label{2}
\end{equation}

At this point, we evoke homogeneity again with the Gibbs-Duheim relation 
\cite{pathria1996statistical} 
\begin{equation}
dG^{\text{field}} = V dP - S^{\text{field}} dT \,.
\end{equation}%
In the bosonic systems treated here, formed by noninteracting massless
fields, $G^{\text{field}}= \mu = 0$. Therefore,
\begin{equation}
dP = \frac{S^{\text{field}}}{V} dT \,.  
\label{3}
\end{equation}
Expression~(\ref{3}) shows that, when the pressure is written in terms of $T$, it does not depend on any other variable, that is $P=P(T)$.

Using relations~(\ref{2}) and (\ref{1}), an explicit form of the
pressure in terms of the entropy can be determined, 
\begin{equation}
S^{\text{field}} = k \mathcal{A} (LT)^{D-1} \, , \,\,
P = \frac{S^{\text{field}}}{V}\frac{T}{k} \,,  
\label{4}
\end{equation}
with a constant $k$. Solving Eq.~(\ref{3}), 
\begin{equation}
dP = \frac{T}{k} \, d\left( \frac{S^{\text{field}}}{V}\right) +
\frac{S^{\text{field}}}{k V} dT = \frac{S^{\text{field}}}{V}dT\,.
\end{equation}
From previous result, 
\begin{equation}
d\left( \ln \frac{S^{\text{field}}}{V T^{D-1}} \right) = 0 
\Longrightarrow 
\frac{S^{\text{field}}}{V} = C T^{D-1} \, ,
\end{equation}
where $C$ is a constant. Finally, using Eq.~(\ref{4}) we obtain 
\begin{equation}
V = \frac{\mathcal{A} \, k}{C} \, L^{D-1} \, .  
\label{volumeV}
\end{equation}

The volume $V$ in Eq.~(\ref{volumeV}) coincides with our choice of effective
volume $V_{\text{eff}}$, up to a nonessential multiplicative constant. We
conclude that the definitions of thermodynamic volume and thermodynamic
limit considered in the present work are compatible with the homogeneity of the equations of state in the thermodynamic limit.

\section{Final remarks}
\label{final-remarks}

In the present work, we considered the equilibrium thermodynamics
associated with massless bosonic fields in anti--de Sitter spacetime with reflexive boundary conditions. Specifically, scalar, electromagnetic and gravitational perturbations are treated. The classical dynamics and the quantization of the several matter contents of interest are based on P\"{o}schl-Teller effective potentials. Thermodynamic quantities are calculated for the fields of interest, and an analysis of the thermodynamics is performed.

It should be remarked that we do not make the usual identification of the cosmological constant $\Lambda$ as proportional to an effective pressure. Although $\Lambda$ is frequently treated as a pressure considering black hole thermodynamics \cite{Kubiznak:2016qmn,Dolan:2011xt}, we consider an alternative formalism compatible with the development presented in Refs.~\cite{Baldiotti:2017ywq,Baldiotti:2016lcf}. In this approach, the thermodynamic interpretation of the cosmological constant is done through  an equation of state. Within this formalism, it is possible to incorporate both the thermal AdS background thermodynamics and the SAdS black hole thermodynamics in the same framework.

An important point in the proposed treatment is a suitable definition of effective volume. This effective volume has a natural geometric interpretation, being compatible with the homogeneity of the equations of state in the thermodynamic limit. Imposing Bose-Einstein statistics and proper thermodynamic limits (based on the volume introduced), the thermodynamics is determined. The proposed definitions of effective volume and AdS thermodynamic limits should be relevant in other analyses in anti--de Sitter background.

In the actual calculation of the partition functions and derived quantities, analytic and numerical tools are employed. Specifically, analytic results are available in the low-temperature and high-temperature regimes, while numeric techniques are (in principle) always applicable. The analytic and numeric calculations coincide when both can be performed, corroborating the approach used.

Bosonic thermodynamics in anti--de Sitter background
is, in many ways, similar to the thermodynamics of more usual physical
systems. For instance, both the AdS bosonic setup and the photon gas in a Minkowskian cavity have the same low-temperature and high-temperature behavior. 
Both scenarios have negative isothermal compressibilities, indicating that they are unstable under isothermic processes. 
Nevertheless, there are distinctions. In the intermediate-temperature regime (with $LT \approx 1$), the AdS thermodynamics is quantitatively different from the ``usual'' boson gas confined in a cavity. In the high-temperature limit, the subdominant terms do not coincide.

It is worth pointing out that, although we considered only bosonic fields, the method presented here can be generalized to other physical scenarios. For instance, fermionic thermal gases in anti--de Sitter are of great interest. Fermions should be part of the Hawking atmosphere in an AdS black hole, and therefore they should be relevant in the Hawking-Page phase transitions. Work along those lines is currently under way.

\begin{acknowledgments}

We thank Alberto Saa, Fernando Brandt, Jorge Noronha and Rodrigo Fresneda for the helpful comments. 
W.~S.~E. acknowledges the support of Coordena\c{c}\~{a}o de Aperfei\c{c}oamento de Pessoal de N\'{\i}vel Superior (CAPES), Finance Code 001, Brazil. 
C.~M. is supported by Grant No.~2015/24380-2, S\~ao Paulo Research Foundation (FAPESP), Brazil; and Grants No.~307709/2015-9 and No.~420878/2016-5, National Council for Scientific and Technological Development (CNPq), Brazil.

\end{acknowledgments}

\end{document}